# Wafer-scale epitaxial growth of single orientation WS$_2$ monolayers on sapphire

*Mikhail.Chubarov[1#], Tanushree H. Choudhury[1#], Danielle Reifsnyder Hickey[1,2], Saiphaneendra Bachu[2], Tianyi Zhang[2], Amritanand Sebastian[3], Anushka Bansal[2], Saptarshi Das[2,3], Mauricio Terrones[2,4,5], Nasim Alem[1,2], and Joan M. Redwing[1,2]*

[1]2D Crystal Consortium-Materials Innovation Platform (2DCC-MIP), Materials Research Institute, The Pennsylvania State University, University Park, PA 16802, USA

[2]Department of Materials Science and Engineering, The Pennsylvania State University, University Park, PA 16802, USA

[3]Department of Engineering Science and Mechanics, The Pennsylvania State University, University Park, PA, USA

[4]Department of Chemistry, The Pennsylvania State University, University Park, PA, USA

[5]Department of Physics, Center for 2-Dimensional and Layered Materials, The Pennsylvania State University, University Park, PA, USA

[#] These authors have contributed equally

*Corresponding authors: tuc21@psu.edu, jmr31@psu.edu

## Abstract

Realization of wafer-scale single-crystal films of transition metal dichalcogenides (TMDs) such as WS$_2$ requires epitaxial growth and coalescence of oriented domains to form a continuous monolayer. The domains must be oriented in the same crystallographic direction on the substrate to avoid the formation of metallic inversion domain boundaries (IDBs) which are a common feature of layered chalcogenides. Here we demonstrate fully-coalesced single orientation WS$_2$ monolayers on 2" diameter c-plane sapphire by metalorganic chemical vapor deposition using a multi-step growth process. High growth temperatures and sulfur/metal ratios were required to reduce domain misorientation and achieve epitaxial WS$_2$ monolayers with low in-plane rotational twist (0.09°). Transmission electron microscopy analysis reveals that the WS$_2$ monolayers lack IDBs but instead have translational boundaries that arise when WS$_2$ domains with slightly off-set lattices merge together. By adjusting the monolayer growth rate, the density of translational boundaries and bilayer coverage were significantly reduced. The preferred orientation of domains is attributed to the presence of steps on the sapphire surface coupled with growth conditions promote surface diffusion and oriented attachment. The transferred WS$_2$ monolayers show neutral and charged exciton emission at 80K with negligible defect-related luminescence. Back-gated WS$_2$ field effect transistors exhibited an $I_{ON}/I_{OFF}$ of ~$10^7$ and mobility of 16 cm$^2$/Vs. The results demonstrate the potential of achieving wafer-scale TMD monolayers free of inversion domains with properties approaching that of exfoliated flakes.

ORCID:
M. Chubarov: 0000-0002-4722-0321
T. H. Choudhury: 0000-0002-0662-2594
S. Bachu: 0000-0001-9898-7349
T. Zhang: 0000-0002-8998-3837
S. Das: 0000-0002-0188-945X
J. M. Redwing: 0000-0002-7906-452X
A. Sebastian: 0000-0003-4558-0013

Transition metal dichalcogenides (TMD) are promising materials for future electronics and optoelectronics due to many attractive and unique properties in the monolayer limit, such as direct bandgap, lack of inversion symmetry and emergence of spin–valley polarization [1–3]. Moreover, the use of TMD materials for electronic devices such as transistors enables reduction of the active channel length compared to that of Si devices, which can be pursued as a route toward further reduction of electronic device size [4,5] and 3D integration [6]. The high photoresponsivity also make

TMDs an interesting choice for photodetectors, photoelectric switches and solar cells [7]. Strained TMD monolayers have been investigated as single-photon emitters with possible applications in quantum photonic devices [8].

Tungsten disulfide ($WS_2$), in the monolayer limit, exhibits a direct bandgap with near bandgap emission at 2 eV [9], has a relatively high mobility [10] and exhibits valley polarization [11]. To achieve the goal of device technologies based on TMDs, large area, single-crystal continuous $WS_2$ monolayers are ideal. Powder vapor transport (PVT), which involves the evaporation of $WO_3$ and S powders in a heated tube furnace, has been successfully employed for the growth of $WS_2$ monolayer domains and films [12,13]. However, the low vapor pressure of $WO_3$ (~0.08 Torr at 1100°C) [14] necessitates the use of furnace temperatures in excess of 900°C to achieve appreciable metal source flux. Additionally, PVT lacks the flexibility to independently control and modulate the source partial pressure during growth.

As a result of these constraints, metalorganic CVD (MOCVD), also referred to as gas source chemical vapor deposition (CVD), has emerged as a promising technique for wafer-scale synthesis of monolayer $WS_2$ as well as other TMD films [15,16]. MOCVD growth of polycrystalline $WS_2$ films was originally reported by Hoffman [17] and later Chung, et al. [18] using tungsten hexacarbonyl ($W(CO)_6$) and hydrogen sulfide ($H_2S$). More recently, organo-chalcogen precursors including diethyl sulfide ($S(C_2H_5)_2$) [15,19] and di-tert-butyl sulfide ($(S(C_4H_9)_2)$ [20] have also been used although simultaneous carbon deposition has been noted in cold-wall reactor geometries [19].

MOCVD growth of coalesced TMD monolayer films on amorphous substrates such as oxidized Si and fused silica leads to randomly oriented domains bounded by high-angle grain boundaries which may serve as scattering centers for charge carriers and are generally undesired. Epitaxial growth on single crystal substrates can potentially eliminate the formation of high-angle grain boundaries and is therefore of significant importance for device-quality TMD films. C-plane sapphire ((0001) α-Al$_2$O$_3$) is a promising substrate for epitaxial growth of WS$_2$ due to its crystallographic compatibility and good thermal and chemical stability. The lattice mismatch between WS$_2$ and sapphire is approximately 30% (a$_{α-Al2O3}$= 4.7597 Å; a$_{WS2}$= 3.1532 Å); nevertheless, the effective lattice mismatch can be significantly reduced assuming domain epitaxy (i.e., three-unit cells of WS$_2$ match with two-unit cells of α-Al$_2$O$_3$). Epitaxial growth of MoS$_2$ on (0001) α-Al$_2$O$_3$ by PVT was initially demonstrated and explained with similar considerations [21]. It was shown that the energetically favorable epitaxial relation is (10$\bar{1}$0) MoS$_2$ ∥ (10$\bar{1}$0) α-Al$_2$O$_3$ [21], which allows the <10$\bar{1}$0> vectors of (0001) α-Al$_2$O$_3$ and MoS$_2$ to be parallel and antiparallel.

In monolayer WS$_2$, as well as in other TMD materials, <10$\bar{1}$0> and <$\bar{1}$010> are not equivalent directions, and domains with opposite directions are referred to as 0° and 60° (or 180°) rotated domains, anti-phase domains or inversion domains. Inversion domain boundaries (IDBs), which form upon coalescence of 0° and 60° domains, have a metallic character which introduces conducting channels in the semiconductor monolayer [22]. IDBs are further problematic for observation of spin and valley polarization which rely on transitions that occur at the K and -K points in the Brillion zone [23]. The presence of inversion domains is readily apparent as oppositely oriented triangles for isolated domains formed on c-plane sapphire as reported for MoS$_2$ grown by PVT [21] and WSe$_2$ [24] grown by MOCVD. However, the presence of steps on the sapphire surface has been reported to induce alignment of isolated WSe$_2$ [25] and MoSe$_2$ [26] domains in PVT growth, providing a possible pathway to reduce or even eliminate IDBs in coalesced epitaxial TMD films.

In this work, we extend these approaches to demonstrate MOCVD growth of epitaxial, fully-coalesced WS$_2$ monolayers on 2" diameter c-plane sapphire that exhibit a preferred crystallographic direction (herein referred to as single orientation) and are largely free of IDBs. A multistep variable temperature growth process combined with the use of high S/W inlet gas phase ratios were necessary to reduce the in-plane rotational domain misorientation, as assessed by in-plane x-ray diffraction, and achieve fully coalesced epitaxial monolayer films. Transmission electron microscopy (TEM) characterization reveals that the $WS_2$ monolayers exhibit a single in-plane orientation and are largely free of IDBs but instead consist of coalesced single crystal regions separated by translational boundaries arising from a slight lattice offset between coalescing domains. The observation of a single orientation across translational boundaries is promising achievement in the direction of achieving single crystal wafer-scale films. The $WS_2$ monolayers transferred to oxidized Si substrates, exhibit clearly resolved exciton, trion and biexciton emission at 80 K with negligible defect-related emission. The electrical characterization of these films shows a $I_{ON}/I_{OFF}$ ratio of $10^7$ at room temperature and promising transfer characteristics. The results pave the way for the realization of wafer-scale epitaxial single-crystal $WS_2$ monolayer films using a process that can be readily extended to other materials in the TMD family.

**Results and Discussion**

In general, high substrate temperatures (>700°C) are required for epitaxial growth of TMDs to provide sufficient thermal energy to promote surface diffusion of metal-containing species [12,21,24]. Under these conditions, however, significant desorption of sulfur will occur due to its high vapor pressure [27], consequently, an excess of chalcogen to metal precursor is required to maintain growth stoichiometry. In this study, the interplay of temperature and sulfur partial pressure was also found to impact the epitaxial orientation of $WS_2$ domains on sapphire. To reveal these effects, a multistep process at constant temperatures of 850 °C or 950 °C was initially studied for growth of $WS_2$ at $H_2S$ partial pressures of 1.7 and 3.4 Torr. The resulting in-plane XRD ϕ-scans of the $\{10\bar{1}0\}$ planes of WS$_2$ are presented in Figure 1(a). A φ-scan of the {30$\bar{3}$0} planes of α-Al$_2$O$_3$ is given for reference. It was observed that the higher H$_2$S partial pressure of 3.4 Torr at higher temperature, as well as lower deposition temperature with H$_2$S partial pressure of 1.7 Torr, both lead to the presence of peaks at only 0° and 60° for the {10$\bar{1}$0} planes of WS$_2$ in Figure 1(a). The epitaxial relation between the substrate and the film is determined to be (10$\bar{1}$0) WS$_2$ ∥ (10$\bar{1}$0) α-Al$_2$O$_3$. Although both conditions lead to epitaxial films, the width of the peak in φ is narrower in the case of the higher deposition temperature (1.13° at 850 °C versus 0.28° at 950 °C). The width of the peak in φ reflects the in-plane rotational twist of the crystal that is associated with edge dislocations and thus can be used as a quality measure of the epitaxial film. WS$_2$ deposited at 950 °C using a lower H$_2$S partial pressure of 1.7 Torr exhibited five crystallographic orientations, characterized by peaks at 0°, 19.1°, 27.5°, 32.5° and 40.9° (Figure 1(a)). The observation that at the deposition temperature of 950 °C, a higher H$_2$S partial pressure led to a reduction in crystallographic orientations demonstrates that the presence of sufficient sulfur on the surface is necessary to promote epitaxial growth. The overall lower intensities of the peaks in the in-plane XRD ϕ-scan (Figure 1(a)) for the film grown at 950 °C and 1.7 Torr suggests that the entire $WS_2$ film is misoriented. If the film consisted of misoriented bilayer or trilayer $WS_2$ domains on a well-oriented underlying monolayer, then the intensity of the peak at ϕ=0° would be close to that of the epitaxial film deposited at 950 °C using the higher $H_2S$ partial pressure (3.4 Torr).

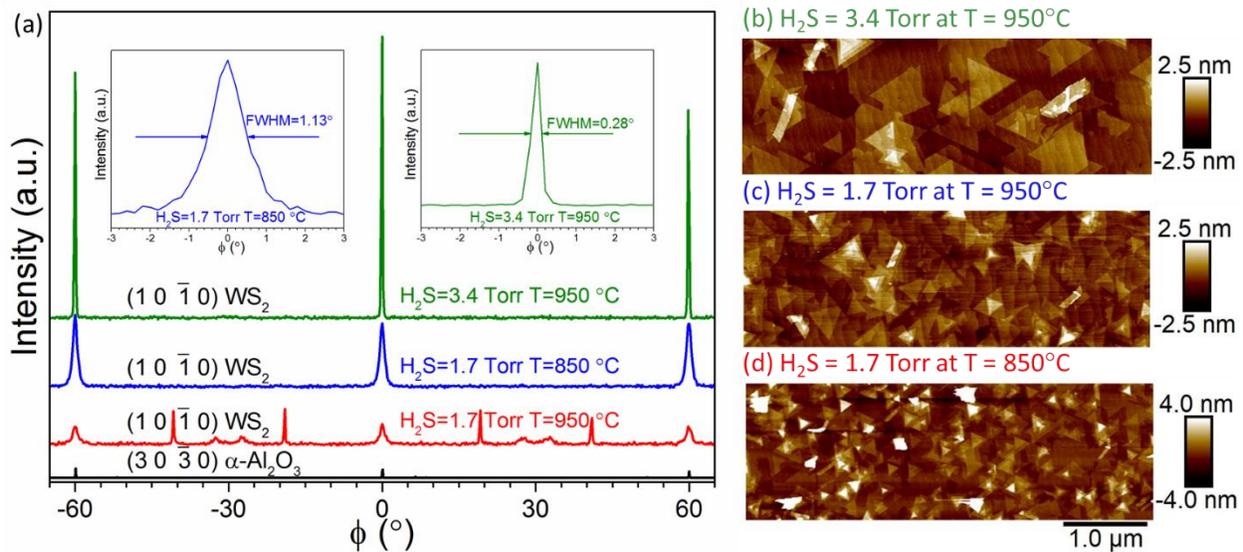

**Figure 1.** (a) In-plane XRD showing φ-scans of {10$\bar{1}$0} planes of WS$_2$ deposited at temperatures of 950 °C and 850 °C and H$_2$S partial pressures of 1.7 and 3.4 Torr. Sapphire substrate {30$\bar{3}$0} φ-scan is presented for reference. Insets show the width of the peak in the φ-scan, which indicates twist of the crystal. Corresponding AFM micrographs of multistep deposition of WS$_2$ (b) 950 °C and 3.4 Torr of H$_2$S, (c) 950 °C and 1.7 Torr of H$_2$S and d) 850 °C and 1.7 Torr of H$_2$S – showing differences in surface morphology

Previously, Ji et al. reported a transition from disordered growth to epitaxy by the introduction of $H_2$ into the growth zone for PVT growth of $WS_2$ using $WO_3$ and S powders as precursors in a hot-wall deposition system [28]. In this case, an increased amount of S was necessary for the deposition of $WS_2$ in the presence of $H_2$ which was attributed to the formation of $H_2S$. The fact that $H_2$ was used as the carrier gas in this study, yet disordered growth was observed at lower $H_2S$ partial pressure, suggests that $H_2S$, rather than $H_2$, must be present in high concentration for single-orientation epitaxy. Additionally, the high $H_2S$ partial pressure is likely necessary to maintain sufficient sulfur-containing species on the surface of the substrate, as the sulfur saturated vapor pressure at 1000 °C is $1.1 \times 10^5$ Torr [27], well in excess of the reactor pressure of 50 Torr.

AFM study of the samples revealed the formation of a continuous $WS_2$ monolayer with a high density of bilayer and multilayer islands (Figure 1 (b-d)). In addition, the surface morphology observed for the sample grown at 950 °C and 1.7 Torr (Figure 1(c)), which consists of a variety of triangular island orientations, agrees well with the in-plane XRD presented in Figure 1(a), in which multiple orientations were observed. Growth at 950 °C with higher $H_2S$ partial pressure (3.4 Torr) also led to larger bilayer domains, as can be seen in Figure 1(b). To combine the benefits of using a lower growth temperature that leads to epitaxy of $WS_2$ and a higher growth temperature that promotes surface diffusion and results in larger domains, a multistep variable temperature process was employed. In this case, $WS_2$ was first nucleated at 850 °C and then annealed in $H_2S$ (4.4 Torr) for 20 minutes at 850 °C. The temperature was then increased to 1000 °C and further annealed (10 minutes duration for this entire step). Next, lateral growth of domains was conducted at 1000 °C with the $W(CO)_6$ concentration in the gas phase reduced to half ($7\times10^{-7}$ Torr) of the nucleation step, whereas the $H_2S$ partial pressure was maintained at 4.4 Torr throughout the process. This, accompanied with a reduction in the growth time, led to a significant reduction in multilayer growth compared to the constant temperature growth at 950 °C represented in Figure 1(b), for which bilayer and trilayer coverages were ~94%, and ~43%, respectively. Under these conditions, a uniform and fully coalesced $WS_2$ monolayer was obtained across the entire 2" sapphire wafer (Figure 2(a)).

The $WS_2$ monolayer follows the morphology of the c-plane sapphire which consists of undulations arising from surface steps (Figure 2(b)). Triangular-shaped bilayer domains are present on the monolayer and range in coverage from ~22% to ~17% from the center to the edge of the wafer, respectively (Figure S1(a, b)). The total growth time (lateral growth) was 45 minutes for this sample, giving an effective growth rate of ~1.6 monolayer/hour (monolayer + bilayer). It should be noted that this growth rate is significantly faster than the growth rates of ~0.04-0.05 monolayer/hour reported for MOCVD $WS_2$ films grown in hot wall reactor configurations using organo-chalcogen sources [12,20]. The use of a lower growth temperature during the nucleation and ripening steps is believed to be beneficial for epitaxy as it provides a higher local S/W ratio on the surface during nucleation to orient the domains and prevents excessive etching/modification of the sapphire during ripening which can occur at higher temperatures. For example, growth of $WS_2$ using the multistep process but at a constant temperature of 1000 °C with otherwise identical conditions resulted in an increase in the surface undulations and non-uniform step structure, which indicates that the $\alpha$-$Al_2O_3$ surface is being modified by high temperature exposure during the nucleation and annealing steps (Figure S1(c)). Additional AFM and SEM images illustrating the surface morphology across an entire 2" wafer are included in Figure S2.

A step height of 0.92 nm was measured across a scratch on the monolayer (Figure 2(b) inset), similar to the value previously reported for single layer $WS_2$ flakes [29]. The step height value of 0.92 nm is larger than the spacing between the {0002} planes of 2H-$WS_2$, which is 0.62 nm [30], and is attributed to the larger van der Waals gap between the film and sapphire substrate, as previously reported for WSe$_2$ on sapphire [31]. Raman spectroscopy using 532 nm and 633 nm laser lines was also used to assess the thickness of the WS$_2$ films on sapphire. However, the Raman peak positions and separation were found to be impacted by residual stress in WS$_2$ on the growth substrate as discussed in supplementary information and Figure S3 and therefore could not be used to accurately assess layer number. Nevertheless, comparison of ultra-low frequency (ULF) Raman spectra of WS$_2$ with different thickness provides further evidence of the monolayer nature of the films as indicated in Figure S4.

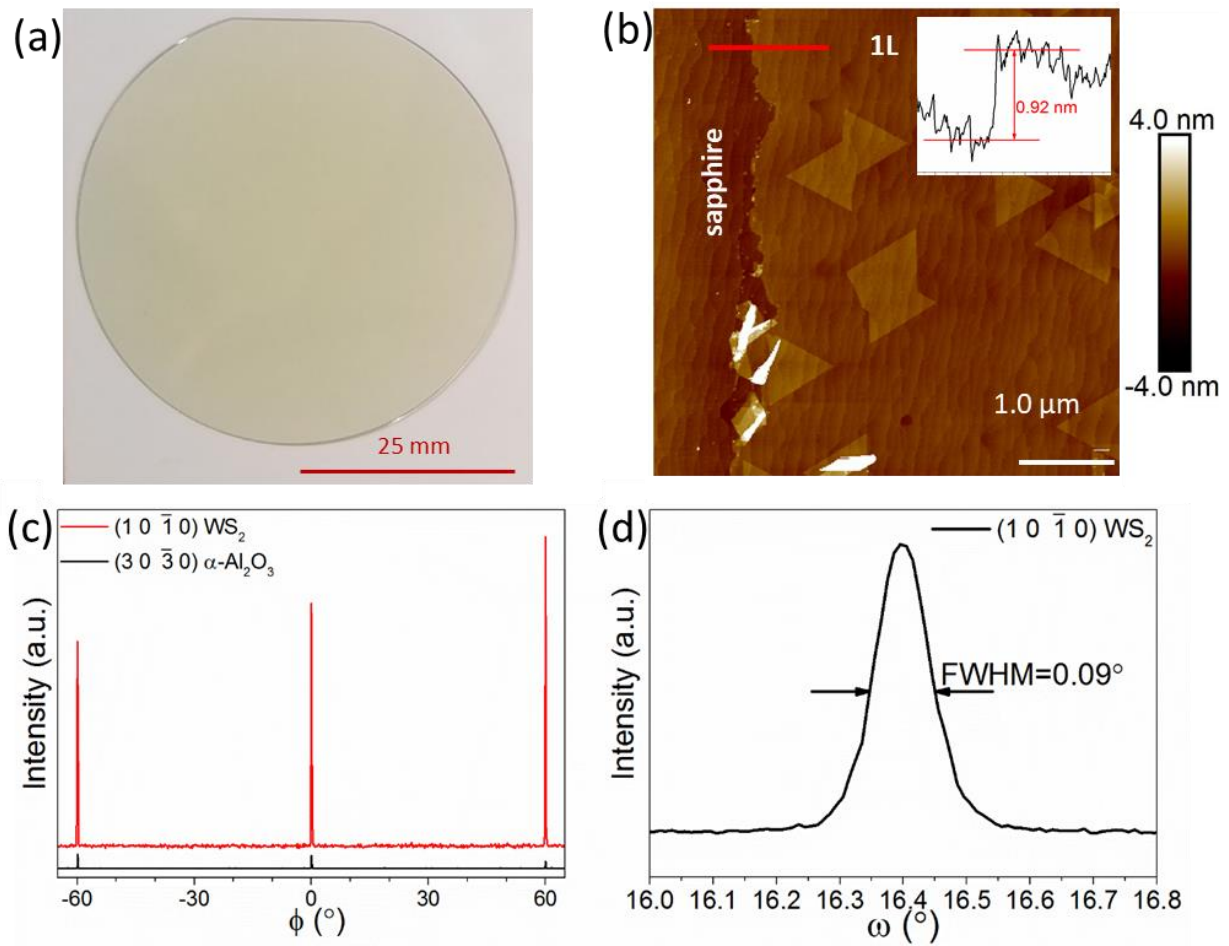

**Figure 2.** (a) Photograph of $WS_2$ monolayer film deposited on two-inch c-plane sapphire substrate using multistep process with varying temperature. (b) AFM topography micrograph of $WS_2$ film showing scratch at the left, monolayer coverage with some areas of bilayer growth. The inset shows the height profile along red line in the top left corner showing height difference of 0.92 nm. In-plane XRD of a $WS_2$ sample deposited using variable temperature multi-step process, showing (c) ϕ-scan of {10$\bar{1}$0} and {30$\bar{3}$0} planes of $WS_2$ and sapphire, indicating epitaxial growth, and (d) ω-scan of {10$\bar{1}$0} planes of $WS_2$, showing low value of in-plane rotational twist.

Figure 2(c) shows in-plane XRD ϕ-scans of the $\{10\bar{1}0\}$ and $\{30\bar{3}0\}$ planes of WS$_2$ and α-Al$_2$O$_3$ respectively. The existence of WS$_2$ peaks separated by 60° from each other and the coincidence of the WS$_2$ peak positions with those of α-Al$_2$O$_3$ indicates that the epitaxial relationship is $(10\bar{1}0)$ WS$_2 \parallel (10\bar{1}0)$ α-Al$_2$O$_3$. Figure 2(d) shows the in-plane XRD ω-scan of the WS$_2$ $(10\bar{1}0)$ plane. The full width at half maximum (FWHM) observed for this peak is 0.09°. This measurement reveals the twist of the WS$_2$ crystal or the misorientation of WS$_2$ domains within the plane of the sapphire. The FWHM value observed here for $(10\bar{1}0)$ WS$_2$ is lower than the FWHM value of 0.28° observed for the film deposited using the multistep process at a constant temperature of 950 °C, presented in Figure 1(a). This suggests that the multistep process with varying temperatures is beneficial for obtaining well-oriented epitaxial WS$_2$ monolayers. It should be noted that the bilayer domains do exhibit a wider distribution of orientations, as shown in Figure S2, likely due to weaker coupling with the sapphire substrate than the monolayer. The in-plane XRD, however, is dominated by the monolayer which has a much higher surface coverage than the bilayer regions.

The WS$_2$ monolayer growth rate was also found to impact the bilayer coverage and microstructure of the film. AFM images of WS$_2$ films deposited using growth rates of 1.6 and 3 monolayer/hour are shown in Figures 3(a) and (b), respectively. At 3 monolayers/hour, the bilayer coverage is further reduced to 0.77% at the center and 0.22% at the edge (Figure S5). Transmission electron microscopy (TEM) was used to study the microstructure of the $WS_2$ monolayers which were removed from the sapphire growth substrates by a water-based transfer method and transferred onto Cu TEM grids (see supporting information). Composite dark-field (DF) TEM images (Figure 3(c) and (d)) were prepared by stitching together a series of DF images such that the structural properties of the $WS_2$ was revealed over an area comparable to that of the AFM scans. The selected area electron diffraction (SAED) pattern corresponding to a single region for both the films has been shown in Figure S6(a, d), while the corresponding DF image is shown in Figure S6(b, e). The SAED patterns of the coalesced $WS_2$ films appear to be single-crystalline: i.e., they contain a single set of spots with the sixfold symmetry of the 2H $WS_2$ crystal lattice. These diffraction patterns represent multi-micron sized areas that have a single, dominant orientation. In the DF-TEM images, the white areas correspond to the bilayer regions. The large circular features present in the composite images are artifacts arising from the holes in the TEM grid. Small black features are also present in the DF-TEM images and are associated with pinholes in the monolayer which may arise from incomplete film coalescence or defects introduced during transfer. The structure of the pinholes is included in Figure S6(g).

Within the $WS_2$ monolayer, two contrasting gray regions (designated region 1 and region 2) are present (Figures 3(c) and (d)). To determine the structure of the two regions and their boundaries, the DF-TEM maps were correlated with atomic-resolution HAADF-STEM imaging. The details of this correlation are discussed in supporting information (Figure S7), the Materials and Methods section and a related publication. It should be noted that these DF-TEM features were observed in every $WS_2$ film studied that was grown using the multi-step process (10 films from different growth runs). Figure 4(a) is a lower magnification dark-field image of an area containing the different regions of contrast. Within a given region, the high-resolution atomic structure is single crystalline (Figure 4(b)). At the interface between two regions Figure 4(c), a line defect is present. Analysis of this defect (Figure 4(d)) reveals no significant angular rotation of the lattice across the boundary, and, furthermore, that the lattice orientation is identical on both sides indicating that this is not an IDB. Instead, the white dashed line in Figure 4(d) shows a translational offset between the two regions separated by the line defect. A schematic of the atomic structure of $WS_2$ has been superimposed on the high-resolution image for ease in visualization. This result is contrary to a previous report of contrast difference in DF-TEM in MoS$_2$ which was attributed to coalescence of 0º and 60º oriented domains [32]. The line defects in the WS$_2$ monolayer arise from coalescing domains that have the same in-plane orientation but lattices that are off-set from one another by less than a unit cell giving rise to translational boundaries. Detailed analysis of the translational boundary structure and geometry indicate that they arise from coalescence of $WS_2$ domains with specific and well-defined edge orientations off the zigzag edges which gives rise to the unique region shapes in the DF images.

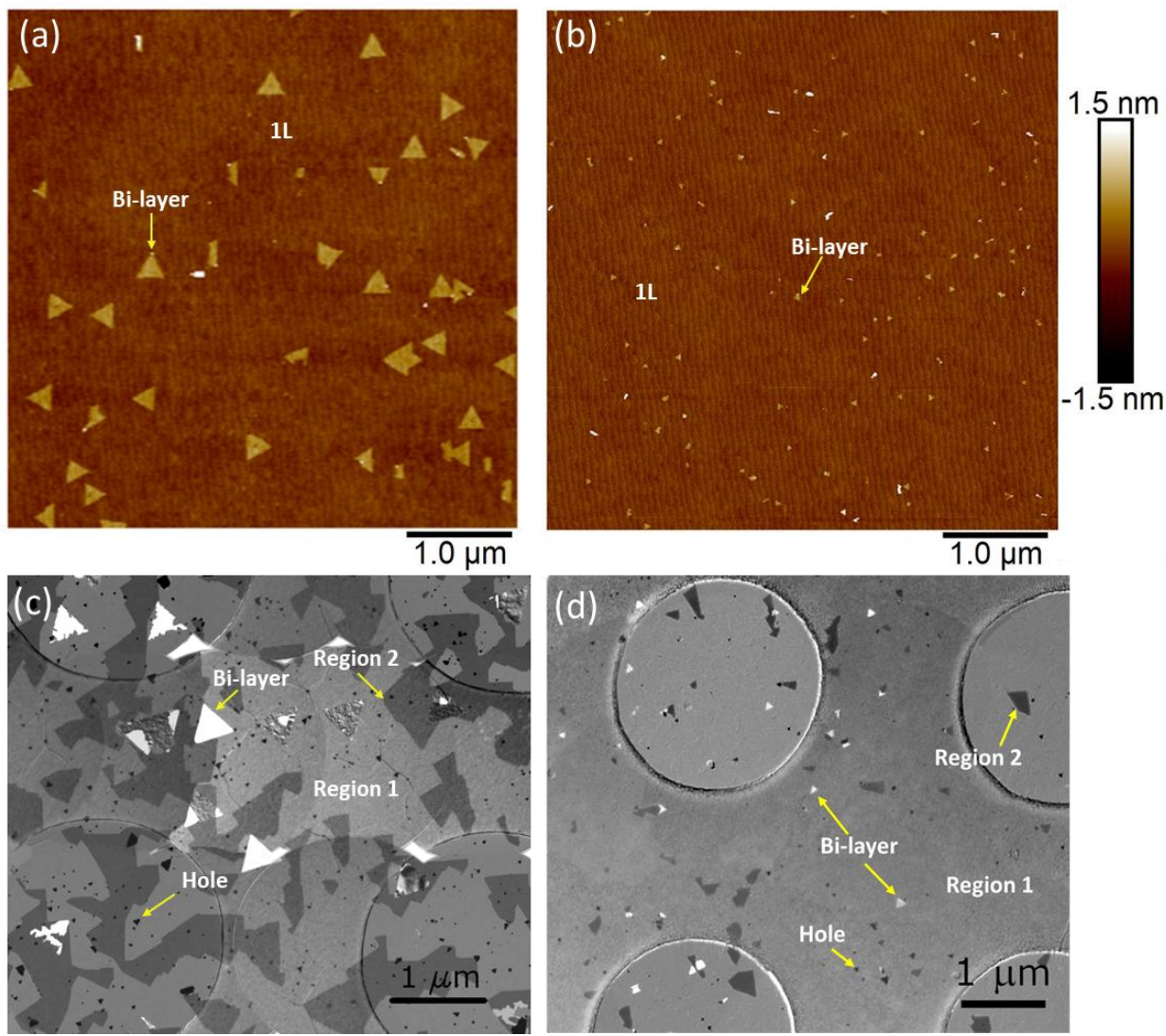

Figure 3. (a) AFM micrograph of films deposited at growth rate of 1.3 monolayer/min (b) AFM micrograph of films deposited at growth rate of 1.3 monolayer/min (c) Composite dark-field TEM map of the film corresponding to (a) and (d) Composite dark-field TEM map of the film corresponding to (b).

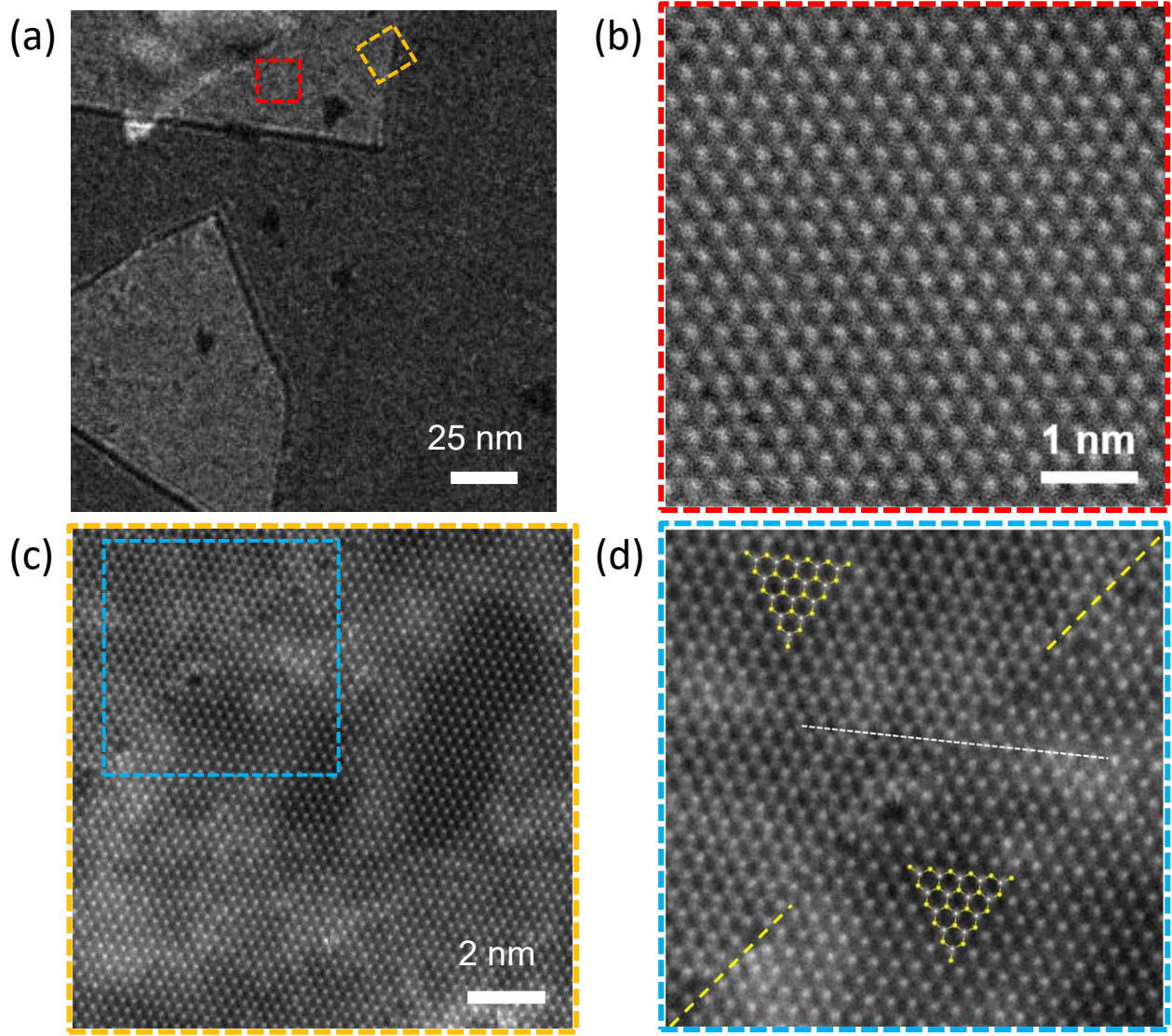

**Figure 4.** (a) A low magnification DF-TEM image showing the two regions. High resolution images showing (b) the 2H WS$_2$ matrix structure (darker gray region in (a)), (c) a region containing the two contrasting areas and (d) higher-resolution atomic structure image showing the line defect separating the two regions of different contrast but same orientation. In (d) the grain boundary is highlighted by a dashed yellow line. A dashed white line across the boundary shows the translational offset between the two regions. The same orientation is confirmed by the relative positions of W and S, which is highlighted by the superimposed WS$_2$ structure.

The contrast in the DF images between regions separated by a translational boundary is the result of differences in the scattering intensity of the sulfur atom columns in the $WS_2$ as can be seen in atomic-resolution HAADF-STEM images (Figures 4(b) and 4(d)). Figure 4(b) shows a region with no grain boundary (acquired from a region of uniform contrast in DF-TEM). Here, the sulfur columns have uniform intensity in the HAADF-STEM image. For comparison, Figure 4(d) shows a region spanning the translational boundary shown in the DF-TEM image in Figure 4(a). Here, the sulfur columns have different intensities. The sulfur columns are clear with bright intensity on the top of the image above the boundary while the sulfur columns are diffused and dimmer in the bottom of the image below the boundary. In atomic-resolution HAADF-STEM imaging, atomic columns appear brighter when they are perfectly on-axis relative to the electron beam. Thus, the region with dimmer sulfur column contrast is slightly off-axis with respect to the beam. This indicates that the region below the boundary is tilted along the z (out-of-plane) axis. This tilt has been previously observed in TMDs and has been reported to occur due to the local strain resulting from the GBs[33]. Therefore, while there is no apparent tilting of the regions in the in-plane direction (within the TMD film), the crystal goes through some degrees of tilt in the out-of-plane direction, changing the Bragg condition locally at the two sides of the grain boundary. This slight tilt leads to different contrasts on the different sides of the GB.

Despite the single orientation in the monolayer, the AFM images in Figure 2 (b) and 3(a) show that the bilayer domains exhibit a variety of orientations including 0° and 60° oriented domains. This variation in bilayer orientation is believed to result from the weaker coupling to sapphire compared to the monolayer and the possibility that some of the bilayer growth may occur at lower temperature during cooldown. Oppositely oriented triangles are commonly observed for epitaxial growth of TMDs on van der Waal surfaces such as graphene since there is no energetic difference between the orientations. Consequently, it is not possible to determine the orientation of the monolayer based on the bilayer orientation as demonstrated in Figure 3(d) where bilayers of opposite orientation are present in single orientation regions of the $WS_2$.

The lack of inversion domains in the $WS_2$ films is surprising but is believed to result from the stepped structure of the sapphire surface and the multistep process employed for layer growth. Figure 5 shows the evolution of the surface morphology of $WS_2$ as a function of time. The c-plane sapphire surface, as received, consists of steps aligned along the $[11\bar{2}0]$ direction with an average terrace spacing of ~70-100 nm (Figure 5(a)). After 30 seconds of nucleation at 850°C and 20 minutes of ripening (850°C-950°C), the surface consists of small ($\leq$10 nm) $WS_2$ clusters that exhibit some alignment relative to one another (Figure 5(b)). During ripening, the clusters are expected to be mobile, similar to that reported for $WSe_x$ clusters on sapphire,[24] and consequently can diffuse to the step edges on the sapphire. The steps may induce a preferred orientation in the $WS_2$ domains similar to observed for $WSe_2$ [25] and $MoSe_2$ [26] grown by PVT. After an additional 6 minutes of lateral growth (Figure 5(c)) triangular domains ~20–50 nm in size are observed which show clustering along some directions. After 10 mins of lateral growth (Figure 5(e)), the domains grow and coalesce and ultimately fill in to form large monolayer regions with a preferred orientation. In Figure 5(e), in-plane XRD φ-scans taken at different times during lateral growth show that the domains exhibit a consistent epitaxial orientation similar to that of the fully-coalesced monolayer (Figure 2(c)). However, the FWHM of the ϕ-scan peaks is considerably larger (~1.0°) at short lateral growth times and decreases as the domains begins to merge (~0.68°) suggesting that the domains become better aligned as coalescence proceeds. The broader ϕ-scan peak at short growth times is consistent with the angular variation of domain orientations (Figure 5(f)) visible in the AFM image (Figure 5(c)) for 6 min growth time. As is clear from Figure 5(f), the domains exhibit a range of orientation angles but the distribution peaks at two angles which are separated by 60°. There is also a clear preference for one orientation over the other, highlighting the preferred orientation in the film even at this early stage of growth.

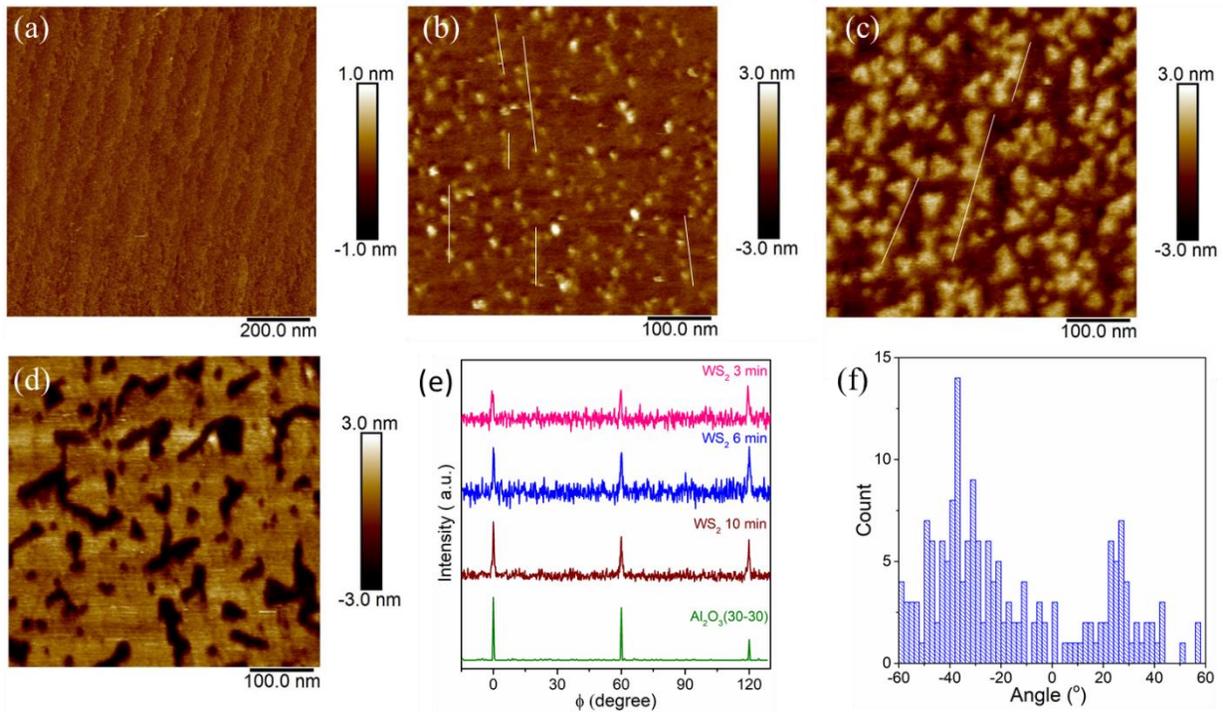

**Figure 5.** AFM micrographs of (a) as-received sapphire substrate, (b) WS$_2$ clusters deposited after nucleation and ripening stage, (c) WS$_2$ monolayer morphology after 6 min. White lines are provided as a guide to the eye. (d) WS$_2$ monolayer morphology after 10 min of lateral growth post nucleation and ripening. (e) In-plane XRD ϕ-scan showing a predominantly single orientation relation between WS$_2$ and underlying α-Al$_2$O$_3$ (f) The orientation distribution observed in panel (c).

An important observation from the AFM images shown in Figure 5(c, d) is the high nucleation density and correspondingly small domain size of the $WS_2$ at the point of coalescence. It is evident that if the final coalesced film microstructure was determined **only** by the domain orientation and size after 6 mins, the monolayer would be interspersed with a large density of grain boundaries separating regions ~100 nm in size, with many small inversion domains. As is evident from Figure 3(c, d), even when the density of line defects is higher, the size of a single-contrast regions (indicative of a single crystalline region enclosed by translational line defects) is of the order of microns. This indicates that small domains near one another merge together seamlessly during coalescence to form a larger single crystalline region. As a result of the lattice mismatch between the $WS_2$ and sapphire, the single crystal regions may be slightly off-set from one another so that when they fuse together a translational boundary like that shown in Figure 4 (c-d) is formed. Evolution from the 6 min microstructure, with a distribution of orientations, to a fully coalesced single orientation film suggests that the small domains can rotate on the sapphire surface and align with nearby domains during coalescence. The steps on the sapphire surface can induce a preferred alignment of domains in one direction giving rise to the single orientation nature of the $WS_2$ monolayers. During lateral growth, the clusters grow in size and merge with one another and nearby domains imparting a preferred crystallographic direction to the coalesced monolayer.

The coalescence process and the resulting microstructure of the monolayer is expected to be dependent on the size and density of the initial $WS_x$ clusters and $WS_2$ domains, the step structure of the sapphire and growth conditions. As shown in Figure 3(c-d), the density of translation boundaries (as measured by the regions of contrast) is dependent on growth rate. At a higher monolayer growth rate, the nucleation density would be higher and hence the $WS_2$ domain size would be smaller enabling migration and alignment of domains during coalescence resulting in larger single crystalline regions. These results demonstrate that large single crystalline regions can be obtained in epitaxial TMD monolayers even when the initial domain density is high as is typically the case for MOCVD and MBE grown material.

The optical and transport properties of fully-coalesced $WS_2$ monolayers were characterized to assess the relative quality of the material. Photoluminescence measurements were obtained over the temperature range from 80 K to 280 K. This is especially useful as the defect-bound exciton intensity was reported to increase as the temperature decreases from 250 K to 77 K [34] indicating that the intensity of defect-bound exciton emission at low temperatures can be used as a measure of film quality. Figure 6 (a) shows that the PL peak for as-grown $WS_2$ on sapphire shifts from 1.95 eV at 280 K to 2.02 eV when the temperature is reduced to 80 K. At 80 K, several spots on the sample were selected to perform PL measurements that showed variation of the peak position between 2.02 eV and 2.04 eV. Note that the low-intensity sharp peak visible at 1.78 eV originates from a Cr impurity in α-$Al_2O_3$ [35]. Negligible defect-bound exciton emission was observed in the low-temperature PL, suggesting that the MOCVD-grown $WS_2$ monolayer has good optical quality. As transferred films are regularly used for device fabrication, low-temperature PL measurements were also carried out on the $WS_2$ monolayer transferred to an $SiO_2$/Si substrate (Figure 6(b)). The 280K PL emission maximum shifted to higher emission energy (2.04 eV), and the FWHM decreased from 82 to 31 meV. The FWHM of the exciton emission reported in the literature for $WS_2$ exfoliated flakes is approximately 30 meV [36,37,38], comparable to the FWHM value observed for the transferred $WS_2$ film. The shift in PL peak position toward higher emission energies upon transfer has previously been observed for PVT-synthesized TMDs films on sapphire, $SiO_2$/Si and fused silica, and was attributed to the relaxation of high-temperature growth-induced strain [39]. The transferred film also does not show prominent emission related to defect-bound excitons, suggesting that the good optical quality of the material is preserved after the transfer. However, additional peaks emerge in the transferred sample at low temperature that can be attributed to trion (highest intensity peak) and biexciton emission, visible as a shoulder on the lower energy side from the trion [37]. The origin of the peaks was determined by excitation laser power dependent measurements as shown in Figure S8. The PL spectra obtained at different powers at 80 K were fitted to extract the peak positions and the power dependence was determined.

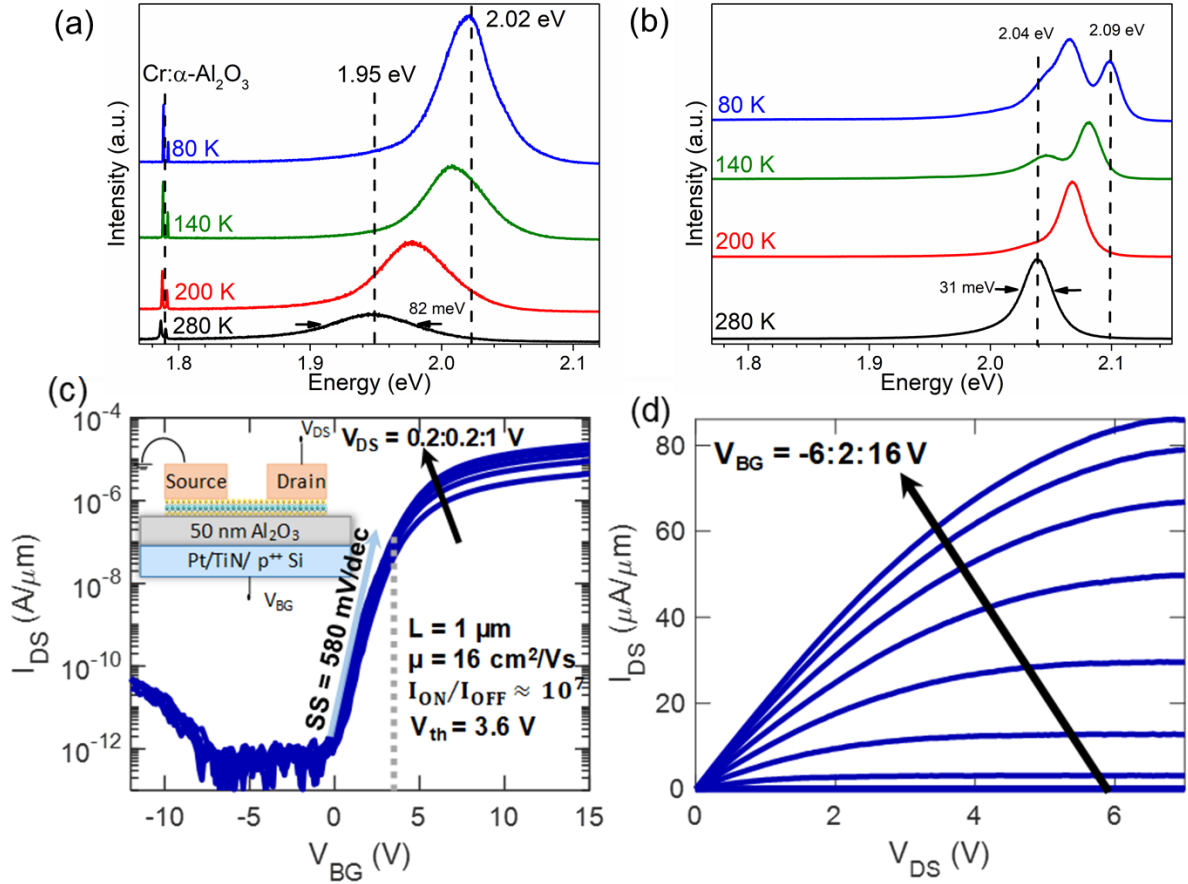

**Figure 6.** Temperature-dependent PL for (a) as-grown WS$_2$ on sapphire substrate and (b) WS$_2$ transferred onto SiO$_2$/Si substrate. (c) Back-gated transfer characteristics for a WS$_2$ FET with the key transistor parameters and the device schematic shown in the inset. The drain current (I$_{DS}$) versus back-gate voltage (V$_{BG}$) for drain voltages (V$_{DS}$) ranging from 0.2 to 1 V is used to find the field effect mobility (μ), subthreshold slope (SS), threshold voltage (V$_{th}$) and ON-OFF ratio ($I_{ON}/I_{OFF}$). (d) The output characteristics of the WS$_2$ FET. The maximum saturation-current is extracted from the output characteristics.

Electrical characterization of the WS$_2$ films was also carried out to assess the transport properties. Back-gated field-effect transistors (FETs) were fabricated (see Electrical Characterization in Materials and Methods Section) after transferring the WS$_2$ film to 50 nm aluminum oxide (Al$_2$O$_3$) with Pt/TiN/p$^{++}$ Si as the substrate and back-gate electrode as shown in the device schematic in the inset of Figure 6(c). Figure S9 show the schematic and the optical image of the device. Figure 6(c) shows the transfer characteristics of a WS$_2$ FET, i.e. the drain current (I$_{DS}$) *versus* the back-gate voltage (V$_{BG}$) for different drain voltages (V$_{DS}$). Clearly, the monolayer WS$_2$ FET shows dominant electron transport along with a weak hole branch consistent with other reports [40,41]. Key FET performance metrics such as the field effect mobility (μ$_{FE}$), sub-threshold slope (SS), threshold voltage (V$_{th}$) and ON-OFF ratio ($I_{ON}/I_{OFF}$) were evaluated. μ$_{FE}$ of 16 cm$^2$/Vs was extracted from the peak transconductance using the equation $\mu_{FE} = g_m L / C_{ox} W V_{DS}$, where $g_m$ is the transconductance $\left(g_m = \frac{\partial I_{DS}}{\partial V_{GS}}\right)$, L is the channel length, $C_{ox}$ is the oxide thickness and W is the width of the channel. V$_{th}$ was found to be 3.6 V using the constant current method (at 100 nA/μm) and SS was found to be 580 mV/decade. A high $I_{ON}/I_{OFF}$ of approximately 10$^7$ is also seen. Figure 6(d) demonstrates the output characteristics, i.e. I$_{DS}$ *versus* V$_{DS}$ for different V$_{BG}$. A high saturation current of 86 μA/μm was measured at a V$_{BG}$ of 16 V and drain voltage (V$_{DS}$) of 7 V corresponding to a carrier density of 1.13 * 10$^{13}$ cm$^{-2}$. The mobilities reported were also comparable to the values in literature. Figure S10 shows the transfer characteristics and statistics of 10 devices fabricated across a 1x1 cm sample region demonstrating a variation of less than +/- 20%. The promising electrical performance further supports the high material quality of the wafer-scale MOCVD grown WS$_2$ film.

## 4. Conclusions

The epitaxial growth of single orientation WS$_2$ monolayer films on 2" c-plane sapphire wafer was demonstrated using a multi-step MOCVD growth process. A transition from multiple preferred orientations of WS$_2$ on the sapphire substrate to epitaxial growth of only one orientation is achieved by increasing the partial pressure of H$_2$S in the reactor. By further employing a variable temperature process and controlling the growth rate, fully-coalesced monolayer WS$_2$ with low in-plane rotational twist (0.09°) and minimal bilayer coverage (<1%) was demonstrated across the entire 2" diameter wafer. TEM characterization of the WS$_2$ monolayer removed from the sapphire reveals micron-size single crystal regions bounded by translational line defects and an absence of inversions domains. The single orientation nature of the WS$_2$ is attributed to the presence of steps on the sapphire surface which serve to induce a preferred alignment and the small size of nucleating domains which may enable oriented attachment. The optical and electrical properties of the WS$_2$ monolayers were characterized for films transferred off the sapphire. Clearly resolved neutral and charged exciton peaks were observed in the photoluminescence spectra obtained at 80K with no prominent defect-related emission. Back-gated WS$_2$ FETs also exhibit a high drive current and a high $I_{ON}/I_{OFF}$. This work demonstrates the possibility of producing wafer-scale *single crystal* TMD monolayers by MOCVD with properties rivaling that of exfoliated flakes.

**Materials and Methods**

**Synthesis.** A cold-wall, horizontal, low-pressure MOCVD reactor was employed for the growth of WS$_2$ films (supporting information Figure S6). Tungsten hexacarbonyl (W(CO)$_6$) and hydrogen sulfide (H$_2$S) were used as the tungsten and sulfur precursors, respectively, with hydrogen (H$_2$) as the carrier gas. W(CO)$_6$ was kept in a stainless-steel bubbler held at a constant temperature (10

°C) and pressure (760 Torr), and $H_2$ gas was passed through it to transport the precursor vapor to the reactor. The inlet gas flow rates of $W(CO)_6$ and $H_2S$ were in the range of $6.4\times10^{-5} – 1.3\times10^{-4}$ sccm ($7\times10^{-7} – 14\times10^{-7}$ Torr) and 160 – 400 sccm (1.7 - 4.4 Torr), respectively. Additional $H_2$ gas was introduced to achieve a total gas flow rate of 4500 sccm through the reactor. Two-inch double-side polished epi-ready c-axis oriented sapphire ((0001) α-$Al_2O_3$) wafers were used as substrates for the growth. No additional treatment was performed on the as-received wafers. The wafers were placed on the SiC-coated graphite susceptor rotating disc. The reactor pressure was kept constant at 50 Torr. Prior to growth, sapphire wafers were held in $H_2$ at a temperature of 850 °C or 950 °C, depending on the growth, for 10 min to remove residual surface contaminants. Growth was initiated by simultaneously introducing the precursors into the inlet gas stream.

For the growth of $WS_2$, a multistep process (nucleation, ripening, lateral growth) was employed similar to that reported by X. Zhang et al. for the growth of $WSe_2$ [24]. Figure S7 in supporting information illustrates the general recipe used for the growth. For the initial experiments, a multistep process at constant temperature was implemented, whereas later experiments were conducted using varying temperatures for the different steps. First, at a temperature of 850 °C or 950 °C, both $H_2S$ and $W(CO)_6$ were introduced for 30 sec with $W(CO)_6$ and $H_2S$ flowrates of $1.3\times10^{-4}$ sccm and 400 sccm, respectively, to form the initial $WS_2$ nuclei on the substrate. In the ripening step, the $W(CO)_6$ was then switched out to bypass the reactor while the flow of $H_2S$ continued into the reactor at a constant rate for 20 min (30 min for initial experiments at constant temperature) to allow surface diffusion and ripening of the $WS_2$ domains to occur. The temperature was then ramped from 850 °C to 1000 °C and held for 10 minutes. After the ripening stage, $W(CO)_6$ was reintroduced into the reactor at half the flow rate ($6.4\times10^{-5}$ sccm) compared to that of the nucleation step for times ranging from 10-45 min to allow the domains to grow laterally in size and coalesce. After growth, the layers were annealed in $H_2S$ flow for an additional 10 minutes at the growth temperature and were then cooled down to 300 °C under a flow of $H_2$ and $H_2S$ to avoid $WS_2$ decomposition. After the temperature decreased below 200 °C, the $H_2$ flow was stopped, the reactor was evacuated to $3\times10^{-3}$ Torr and a reactor purging procedure using $N_2$ was initiated to remove any residual $H_2S$ from the reactor before the sample was removed. The sample was then unloaded from the reactor into a $N_2$-purged glovebox to avoid exposing the sample and reactor directly to atmospheric moisture and oxygen. Between characterization studies, the samples were stored in a $N_2$-ventilated cabinet to reduce further sample degradation.

**Film transfer details.** The films were transferred from sapphire for transmission electron microscopy (TEM), low temperature photoluminescence (PL) and electrical characterization. The typical transfer procedure involves coating an as-grown sample with polymethametaacrylate (PMMA) using a spin-coater in two steps: 500 rpm for 15 seconds followed by 4500 rpm for 45 seconds. After the PMMA coating cures overnight, the sample edges are scratched to assist delamination, and the sample is then immersed into 1M NaOH solution in DI water and held at 90 °C for 15-20 minutes. NaOH helps delaminate the $WS_2$ film from the substrate, and when it does, the PMMA+$WS_2$ film assembly floats on the surface of the NaOH solution. The assembly is then transferred to a DI water bath for 10 minutes for rinsing. This step is then repeated three more times to ensure the complete removal of residual NaOH from the previous step. The assembly is then fished out using a 3 mm diameter Cu Quantifoil TEM grid, $SiO_2$/Si and 50 nm $Al_2O_3$ with Pt/TiN/p$^{++}$Si as the back-electrode stack for TEM, PL and electrical characterization respectively. In the case of TEM, transfer is done so that the Quantifoil side touches the $WS_2$ film. The PMMA+$WS_2$ film + TEM grid assembly is heated at 50 °C for 10 minutes and at 70 °C for 10 more minutes before placing it into an acetone bath for the removal of PMMA film. The TEM grid, which has the $WS_2$ film on the Quantifoil side, is then transferred into an alcohol (methanol or isopropanol) bath to remove acetone residue. Finally, the grid is heated at 70 °C on a hotplate for 10 minutes. The same procedure is followed for the transfer of the films to the other substrates.

**Material characterization.** A Bruker Icon atomic force microscope (AFM) was used to study the surface morphology, domain size, coverage and thickness of the deposited layers. Scanasyst air probe AFM tips with a nominal tip radius of ~2 nm and spring constant of 0.4 N/m were employed for the measurements, and images were collected using peak-force tapping mode. To measure the thickness of the deposited films, samples were lightly scratched using a blunt tweezer to remove a portion of the weakly bonded $WS_2$ film without damaging or scratching the sapphire surface.

A Zeiss Merlin electron microscope was used for acquiring the scanning electron micrographs. An accelerating voltage of 3 kV and a working distance of 3 mm was used with the inlense detector to capture the images.

A PANalytical MRD diffractometer with 5-axis cradle was employed for X-ray diffraction characterization of $WS_2$ films. A standard Cu anode X-ray tube operated at 40 kV accelerating voltage and 45 mA filament current was used to generate X-rays. As primary optics, a mirror with ¼° slit and Ni filter was used to discriminate the Cu K$_\beta$ line. On the diffracted beam side, an 0.27° parallel plate collimator with 0.04 rad Soller slits with PIXcell detector in open detector mode were employed. Samples were positioned in such a way that the sample surface was ~2-4° away from the X-ray incidence plane. Such a configuration allows measurement of diffraction caused by the (hki0) planes to determine the in-plane epitaxial relation of the film with respect to a substrate, as previously reported [42].

Raman measurements were performed using a HORIBA LabRAM HR Evolution Raman microscope with laser wavelengths of 532 nm and 633 nm. For Raman measurements, a grating with 1800 groves per mm was employed. Raman $WS_2$ signature positions as well as the position and intensity of the near bandgap emission in PL from $WS_2$ were used to confirm the formation of mainly monolayer thin films.

Temperature-dependent PL measurements were performed under 488 nm laser excitation using a Renishaw inVia Raman microscope. A Linkam THMS600 optical stage was used as the sample holder during PL spectra acquisition. All measurements were performed in nitrogen atmosphere.

The deposited $WS_2$ films were imaged using various TEM techniques to reveal the structure of the films at the atomic level. High-angle annular dark-field scanning TEM (HAADF-STEM) imaging was performed on a FEI Titan3 G2 operating at an accelerating voltage of 80 kV with a probe convergence angle of 30 mrad and probe current of 70 pA. Dark-field (DF-) TEM imaging was performed on a FEI Talos F200X using a 10 μm objective aperture to select a $\{10\bar{1}0\}$ diffraction spot. DF-TEM images were acquired for 30 s each. Composite DF-TEM images were prepared for the $WS_2$ to reveal the grain distribution on a larger scale. The composite DF-TEM map is built from individual micrographs using GIMP 2 image processing software to adjust the contrast between different micrographs and within a single micrograph, in which areas with free-standing film are accompanied by areas where the film is supported by the carbon TEM grid. The composite image preparation used here is similar to that reported by P. Y. Huang et al. for graphene [43].

**Process to correlate DF-TEM features with high resolution images.** Boundaries were first located in atomic-resolution imaging in order to image the films in the cleanest state, as free of hydrocarbon-based contamination as possible. A series of HAADF-STEM images were collected while moving along each GB to determine the atomic structure. Then, the same regions were imaged at low magnification in DF-TEM to correlate the atomic-resolution structure with the DF-TEM patterns. Although the monolayer films are very uniform, imperfections such as small holes in the film or small bilayer regions were used as landmarks to be certain that the regions examined at different magnifications were exactly correlated. Due to the fact that the film is freestanding on the carbon Quantifoil TEM support, neither grain is exactly aligned along the c axis relative to the electron beam, and strain and tilting in the free-standing films caused the sulfur column intensity on one side of the GB to be higher than on the other. Although sulfur atoms (Z=16) have much lower HAADF-STEM intensity than tungsten atoms (Z=74) (HAADF-STEM image intensity is proportional to $\sim Z^{1.8}$), it was possible to unambiguously identify that the $WS_2$ film has the 2H crystal structure on both sides of the GB, with the same x-y orientation. Despite the 2H lattice having the same orientation on both sides of these GBs, a translational shift was observed between the grains, with no rotational misalignment. NB, direct HAADF-STEM imaging across these boundaries also ruled out 60°-rotated mirror-twinned regions in the regions studied.

**Electrical characterization.** The as-grown $WS_2$ on sapphire substrate is transferred to a substrate with 50 nm $Al_2O_3$ with Pt/TiN/p$^{++}$Si as the back-electrode stack. Electron beam (e-beam) lithography (Vistec EBPG5200) is used for the isolation step to define the channel dimensions. After the electron-beam exposure and the develop step, $SF_6$ etch (PT Dual Etch Versalock) is done to define the device area. Then the source and drain electrodes are defined using another e-beam lithography step to obtain a channel length (L) of 1 μm. 40 nm Ni/30 nm Au stack is deposited through e-beam evaporation (Temascale FC2000) to obtain the final back-gated field effect transistor (FET) device structure. Following the device fabrication, the electrical characterization was carried out at room temperature in high vacuum ($\approx 10^{-5}$ Torr) in a Lakeshore CRX-VF probe station using a Keysight B1500A parameter analyzer.

**Acknowledgements**

Primary financial support for this work was provided by the National Science Foundation (NSF) through the Pennsylvania State University 2D Crystal Consortium–Materials Innovation Platform (2DCC-MIP) under NSF cooperative agreement DMR-1539916. D.R. Hickey., S. Bachu. and N. Alem. acknowledge additional support from an NSF CAREER grant (DMR-1654107). T. Zhang. and M. Terrones. acknowledge the financial support of the Air Force Office of Scientific Research (AFOSR) through grant FA9550-18-1-0072. The work of S. Das. and A. Sebastian. was partially supported by Army Research Office through Cooperative Agreement Grant W911NF1920338 and Air Force Office of Scientific Research (AFOSR) through the Young Investigator Program Grant FA9550-17-1-0018.

*Supporting Information for*

# Wafer-scale epitaxial growth of single-orientation WS$_2$ monolayers on sapphire

*Mikhail.Chubarov[1#], Tanushree H. Choudhury[1#], Danielle Reifsnyder Hickey[1,2], Saiphaneendra Bachu[2], Tianyi Zhang[2], Amritanand Sebastian[3], Saptarshi Das[2,3], Mauricio Terrones[2,4,5], Nasim Alem[1,2], and Joan M. Redwing[1,2]*

[1]2D Crystal Consortium-Materials Innovation Platform (2DCC-MIP), Materials Research Institute, The Pennsylvania State University, University Park, PA 16802, USA

[2]Department of Materials Science and Engineering, The Pennsylvania State University, University Park, PA 16802, USA

[3]Department of Engineering Science and Mechanics, The Pennsylvania State University, University Park, PA, USA

[4]Department of Chemistry, The Pennsylvania State University, University Park, PA, USA

[5]Department of Physics, Center for 2-Dimensional and Layered Materials, The Pennsylvania State University, University Park, PA, USA

[#] These authors have contributed equally

*Corresponding authors: tuc21@psu.edu, jmr31@psu.edu

ORCID:
M. Chubarov: 0000-0002-4722-0321
T. H. Choudhury: 0000-0002-0662-2594
S. Bachu: 0000-0001-9898-7349
T. Zhang: 0000-0002-8998-3837
S. Das: 0000-0002-0188-945X
J. M. Redwing: 0000-0002-7906-452X
A. Sebastian: 0000-0003-4558-0013

**AFM images of WS$_2$ on sapphire corresponding to Figure 2 and Figure 3 (a, c).**

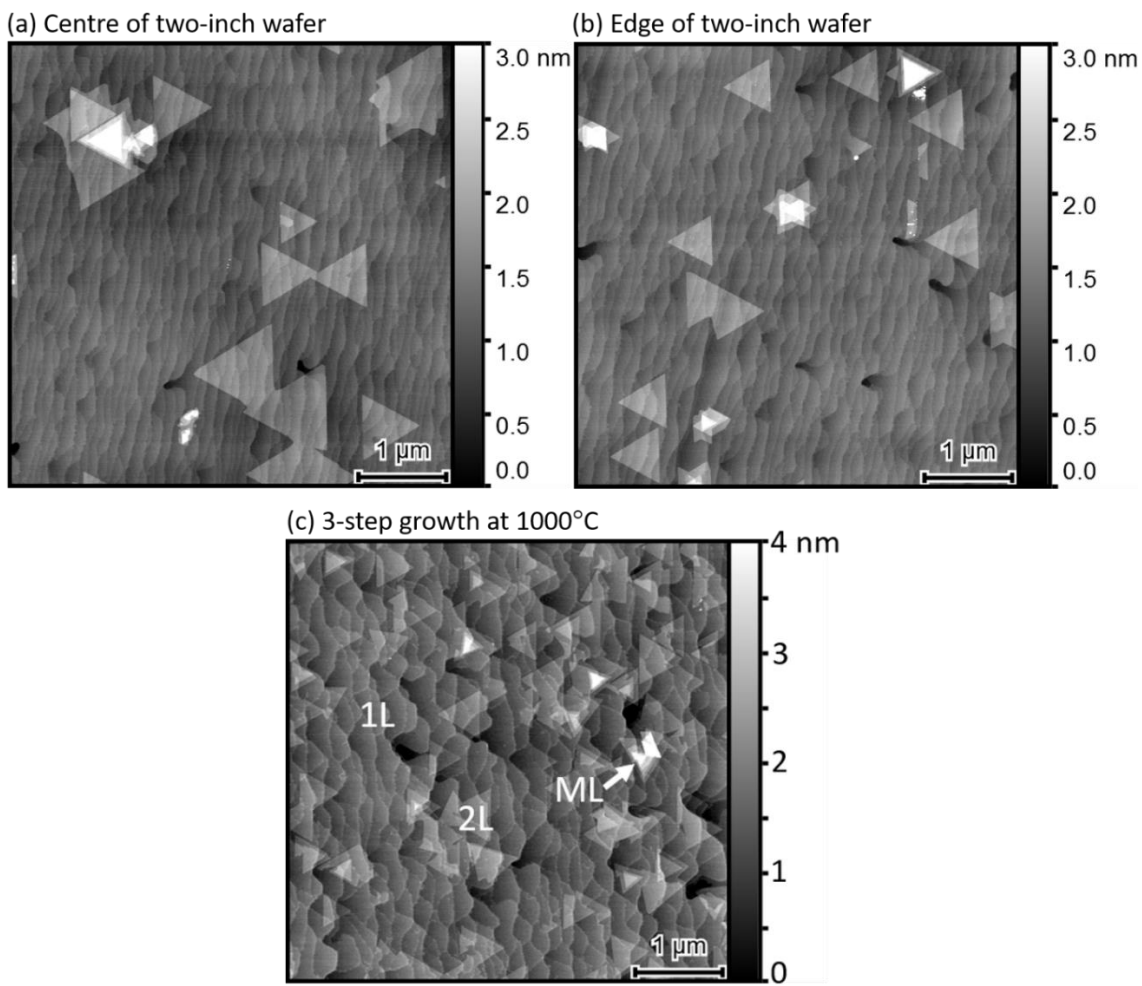

**Figure S1.** AFM micrographs recorded a) in the center and b) at the edge of WS$_2$ film deposited on two-inch sapphire wafer using three-step process with nucleation at 850 °C and growth at 1000 °C, showing a monolayer film (1L) with some bilayer (2L) regions. (c) AFM micrograph of WS$_2$ film deposited using three-step process with nucleation, ripening and growth at 1000 °C showing coarsening of the sapphire surface.

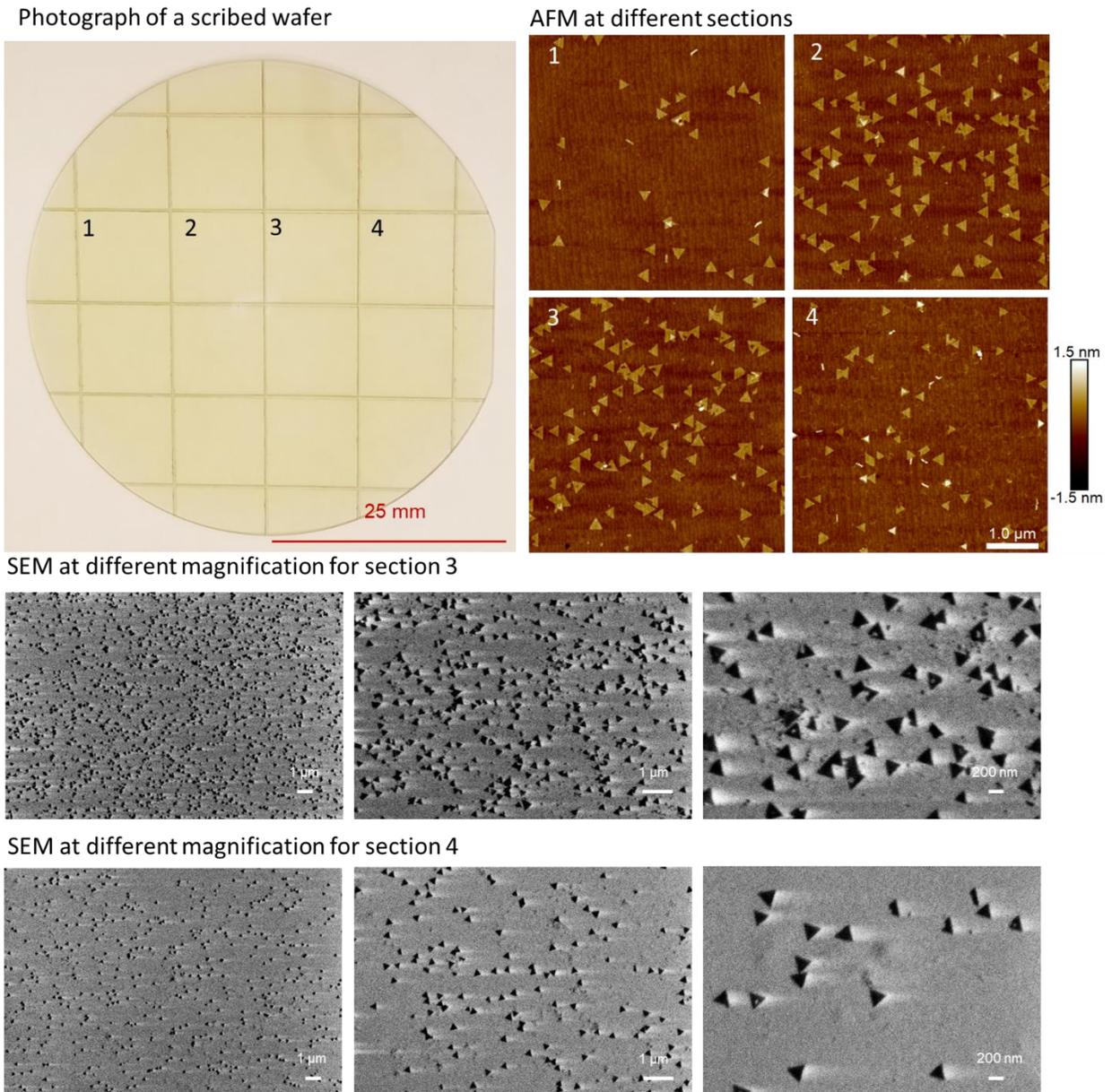

**Figure S2.** Photograph of $WS_2$ on a 2" sapphire wafer (with scribe marks on the backside) and AFM micrographs from designated sections of the wafer. SEM micrographs at different magnifications from several sections of the wafer are included to highlight the uniformity at difference length scales. The bilayer density varies from ~12 % at the center to ~ 5 % at the edge.

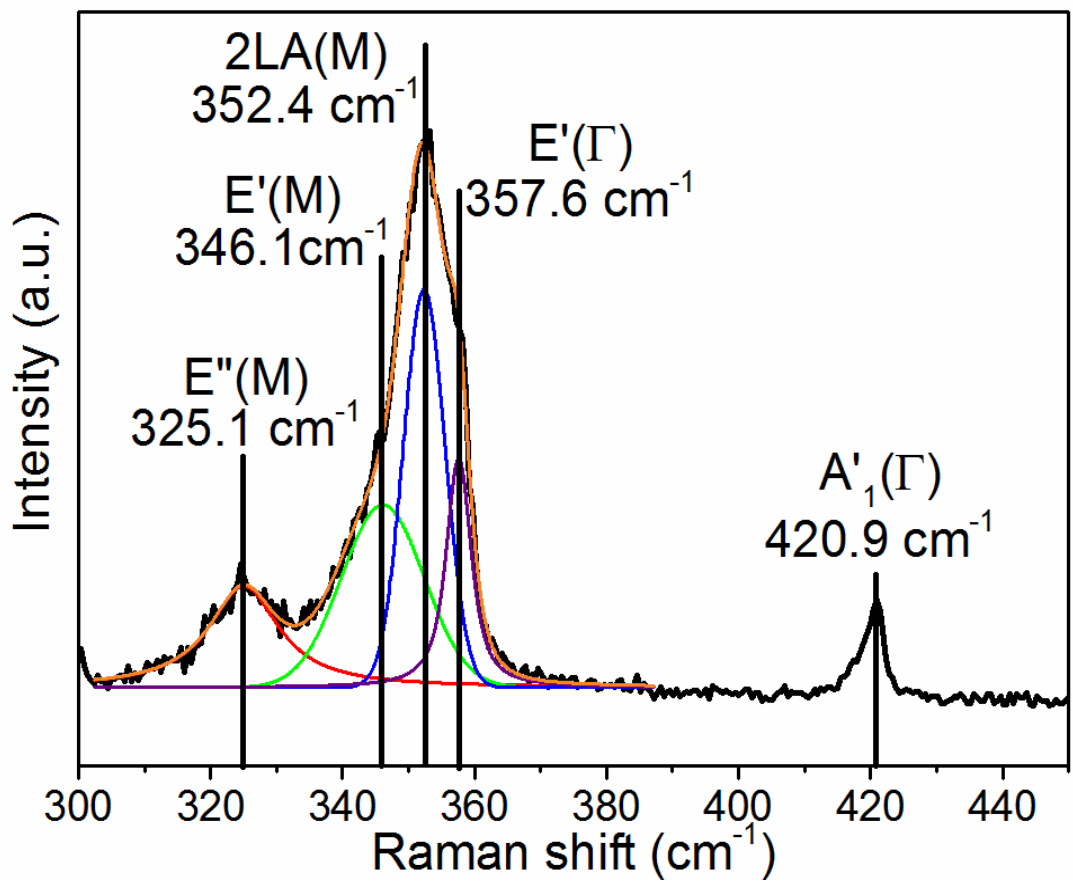

**Figure S3**. Raman spectra recorded for monolayer $WS_2$ as deposited on c-plane sapphire substrate. Figure shows shape of the spectra characteristic for the monolayer film while spacing between E'(Γ) and $A_1$'(Γ) corresponds to 3-4 layer film.

**Raman spectroscopy measurements of layer thickness**. The use of Raman spectroscopy to determine of the number of layers in the $WS_2$ film on sapphire based on the spacing between Raman modes was complicated due to the presence of residual stress in the films. Comparison of the spectrum shape in the range between 270 and 480 cm$^{-1}$ to what has been reported by other researchers and, for example, by M. O'Brian et al. [1] has been used to identify the formation of a monolayer film due to the considerably higher intensity of the peak at approximately 352 cm$^{-1}$ compared to the $A_1$'(Γ) at approximately 420 cm$^{-1}$. Such behavior has been observed from the samples deposited in this work (Figure S2).

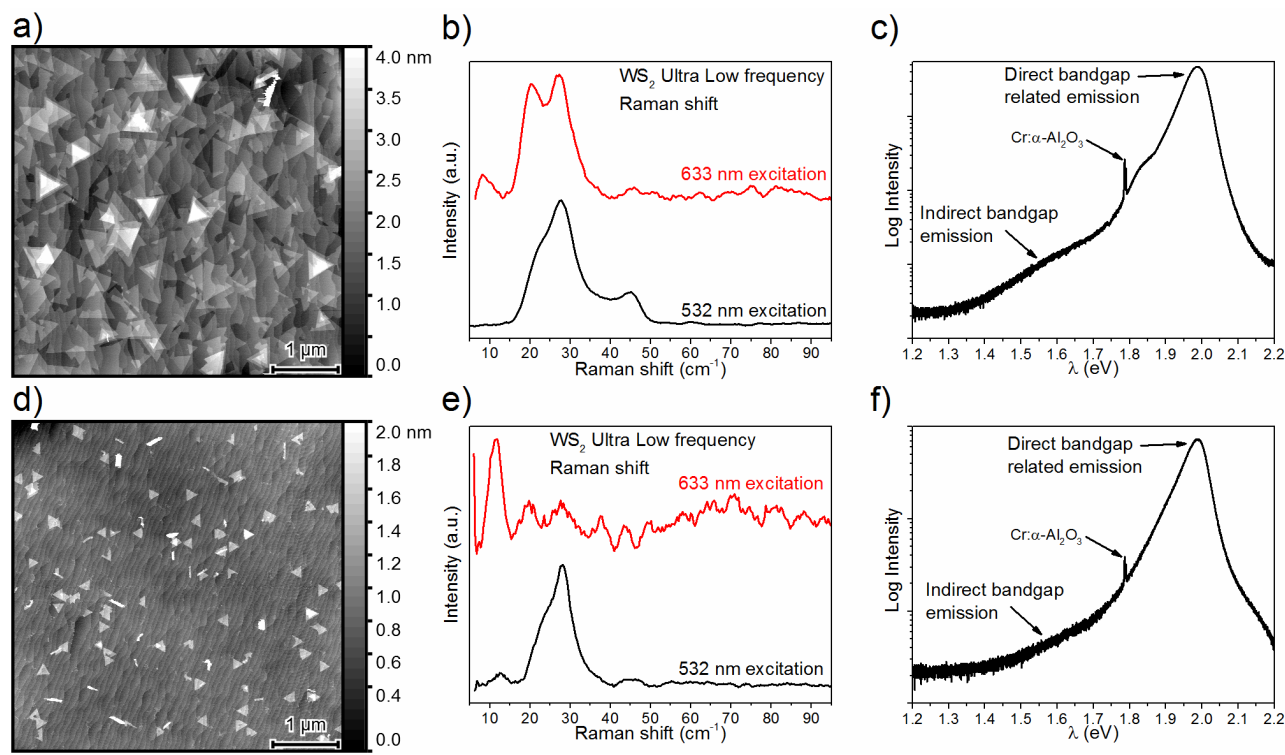

**Figure S4.** a) AFM image of multilayer $WS_2$ film, b) ULF Raman spectra recorded using 633 nm laser (upper red curve) and 532 nm laser (bottom black curve) for $WS_2$ film in figure a), c) PL spectra recorded using 532 nm laser for $WS_2$ film shown in figure a). d) AFM image of primarily monolayer $WS_2$, e) ULF Raman spectra recorded using 633 nm laser (upper red curve) and 532 nm laser (bottom black curve) for $WS_2$ film in figure d), f) PL spectra recorded using 532 nm laser for $WS_2$ film shown in figure d). For the better visualization of the ULF Raman spectrum, baseline was subtracted from the raw data. Noise visible for the non-resonant ULF Raman in b) and e) is an effect likely caused by background removal process and is not expected to be related to low frequency modes of $WS_2$.

**Ultra-low frequency (ULF) Raman measurements** were successfully implemented as a better indication of the number of layers; however, it has been reported by M. O'Brian et al. that using resonant Raman measurements, a peak at 27 cm$^{-1}$ is observed for the monolayer films and can be related to the LA(M) resonant peak [1]. This complicates the determination of number of layers in the film. To resolve this, we deposited two $WS_2$ samples where one exhibits a multilayer structure and the other is primarily a monolayer film (Figure S4 (a) and (d)). The baseline-subtracted ULF Raman spectra recorded using 633 and 532 nm lasers are presented in Figure S4 (b) and (e). From these experiments, we observe the presence of peaks in the ULF Raman spectrum for a multilayer film for both the 633 and 532 nm excitation lasers, whereas for the monolayer films, a well-defined peak was only observed for the 532 nm laser. This observation further supports the results reported by O'Brian et al.[1]. Finally, a plot of the PL spectra for the two samples mentioned above reveals a low-intensity feature at approximately 1.77 eV for the multilayer film that has considerably lower intensity for the primarily monolayer film, and the likely origin for this feature is fluorescence from the substrate (Figure S3 (c) and (f)). It is worth noting that the intensity of the indirect transition-related emission at 1.77 eV is of very low intensity as both graphs represent logarithmic intensity (Figure S3 (c) and (f)).

**AFM images of WS$_2$ on sapphire corresponding to Figure 3 (b, d)**

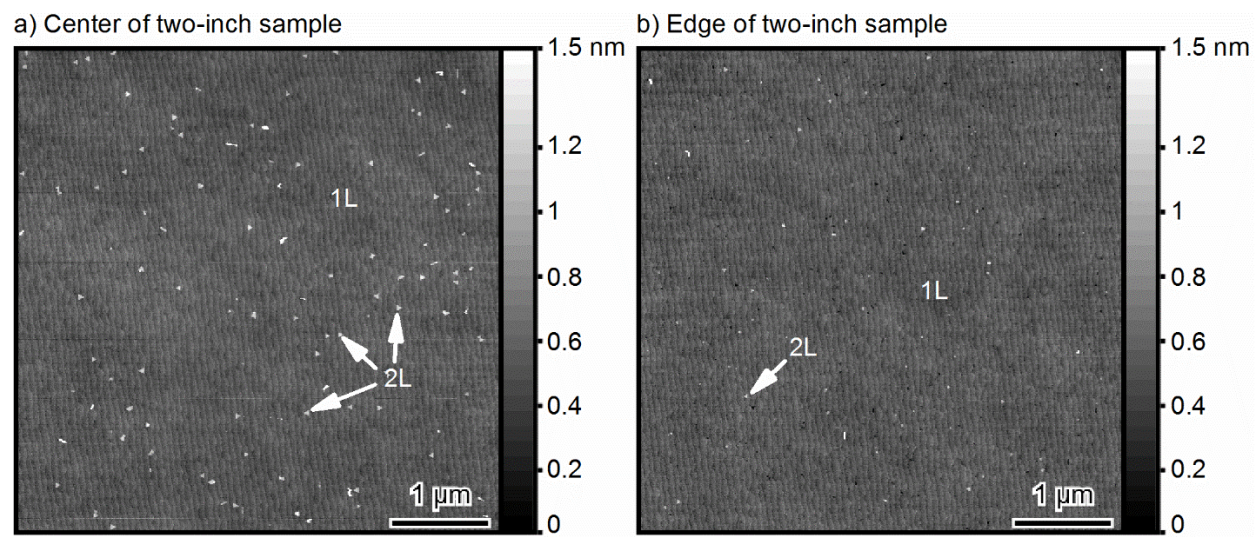

**Figure S5.** AFM micrographs recorded a) in the center and b) at the edge of WS$_2$ film deposited on two-inch sapphire wafer using three-step process with nucleation at 850 °C and growth at 1000 °C at a growth rate of 3 monolayer/hour, showing a monolayer film (1L) with bilayer (2L) regions visible as small, bright spots.

**SAED Pattern and additional TEM images corresponding to Figure 3.**

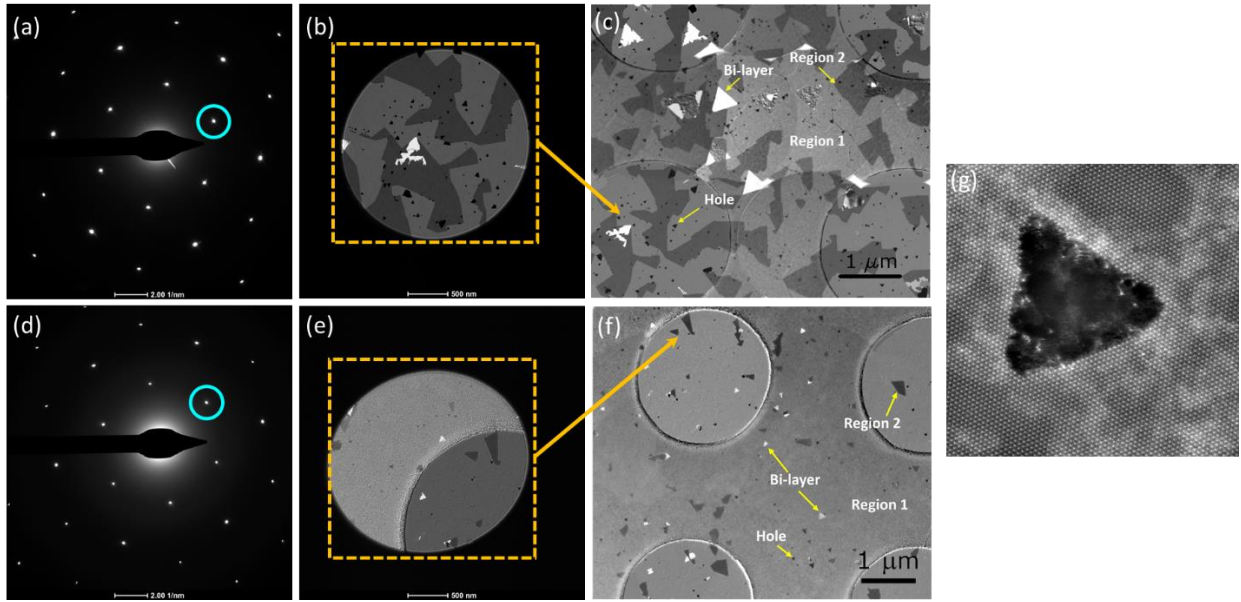

**Figure S6.** (a) SAED pattern for the WS$_2$ film region shown in (b) from the DF-TEM map in Figure. 3(c). The DF-TEM image was formed by selecting the area shown in cyan with a 10 μm objective aperture. (c) Figure 3(c) is included to show the region from which the SAED and DF-TEM image were captured. (d) SAED pattern for the WS$_2$ film region shown in (e) from the DF-TEM map in Figure 3(d). The DF-TEM image was formed by selecting the area shown in cyan with a 10 μm objective aperture. (f) Figure 3(d) is included to show the region from which the SAED and DF-TEM image were captured. In the Figure S6(e) region, the film is partly freestanding and partly suspended over carbon TEM grid support). (g) Structure of a triangular pinhole present in the films resulting from incomplete coalescence.

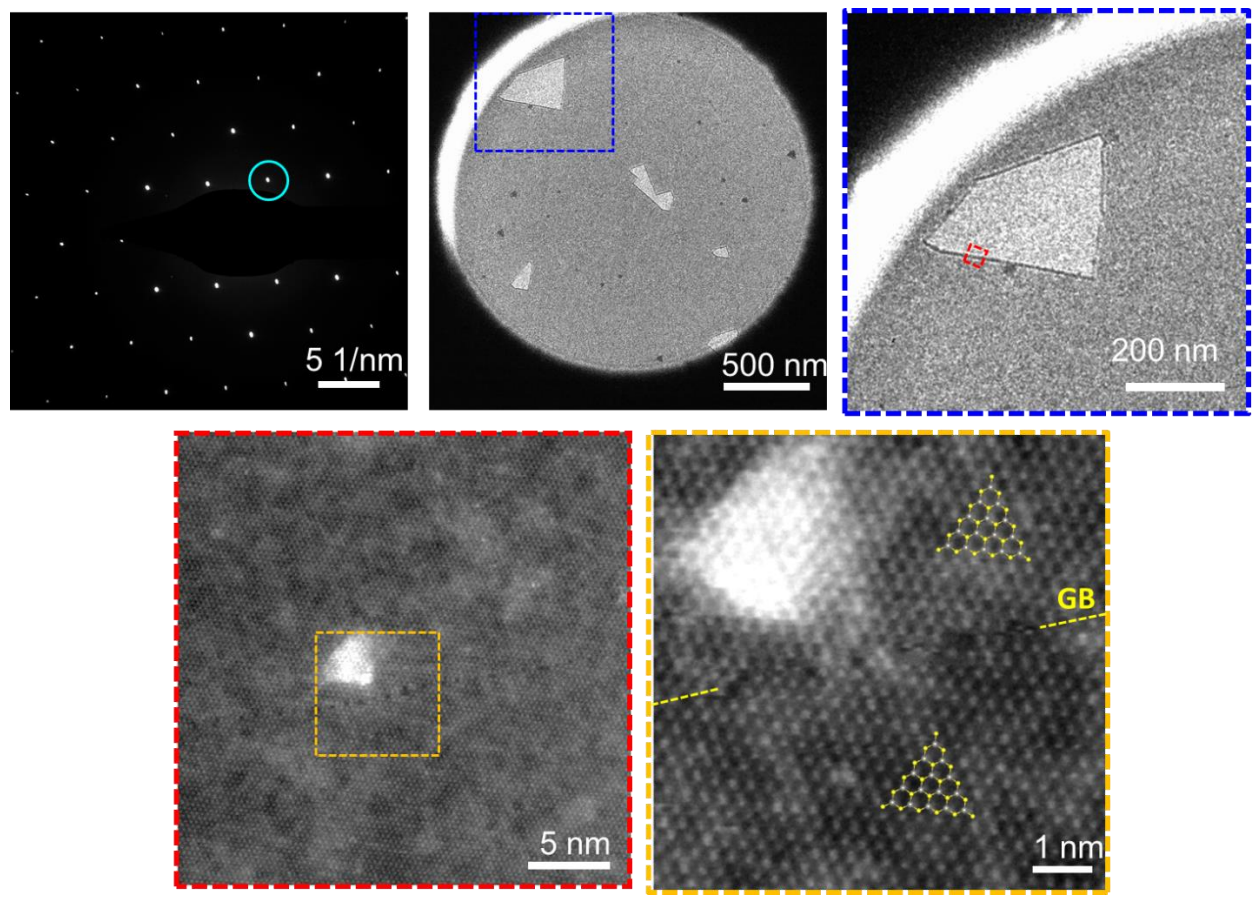

**Figure S7.** (a) SAED pattern and (b) DF-TEM collected from cyan circle. (c) Blue dashed region is zoomed in and shows red square indicating position of HAADF-STEM imaging. (d) HAADF-STEM imaging shows translational grain boundary with a bright spot (atomic cluster) used as a reference point for imaging. (e) In the highest-magnification HAADF-STEM image, the same orientation of the 2H lattice can be observed on opposite sides of the grain boundary as highlighted by the superimposed $WS_2$ structure.

**Power dependent PL analysis**

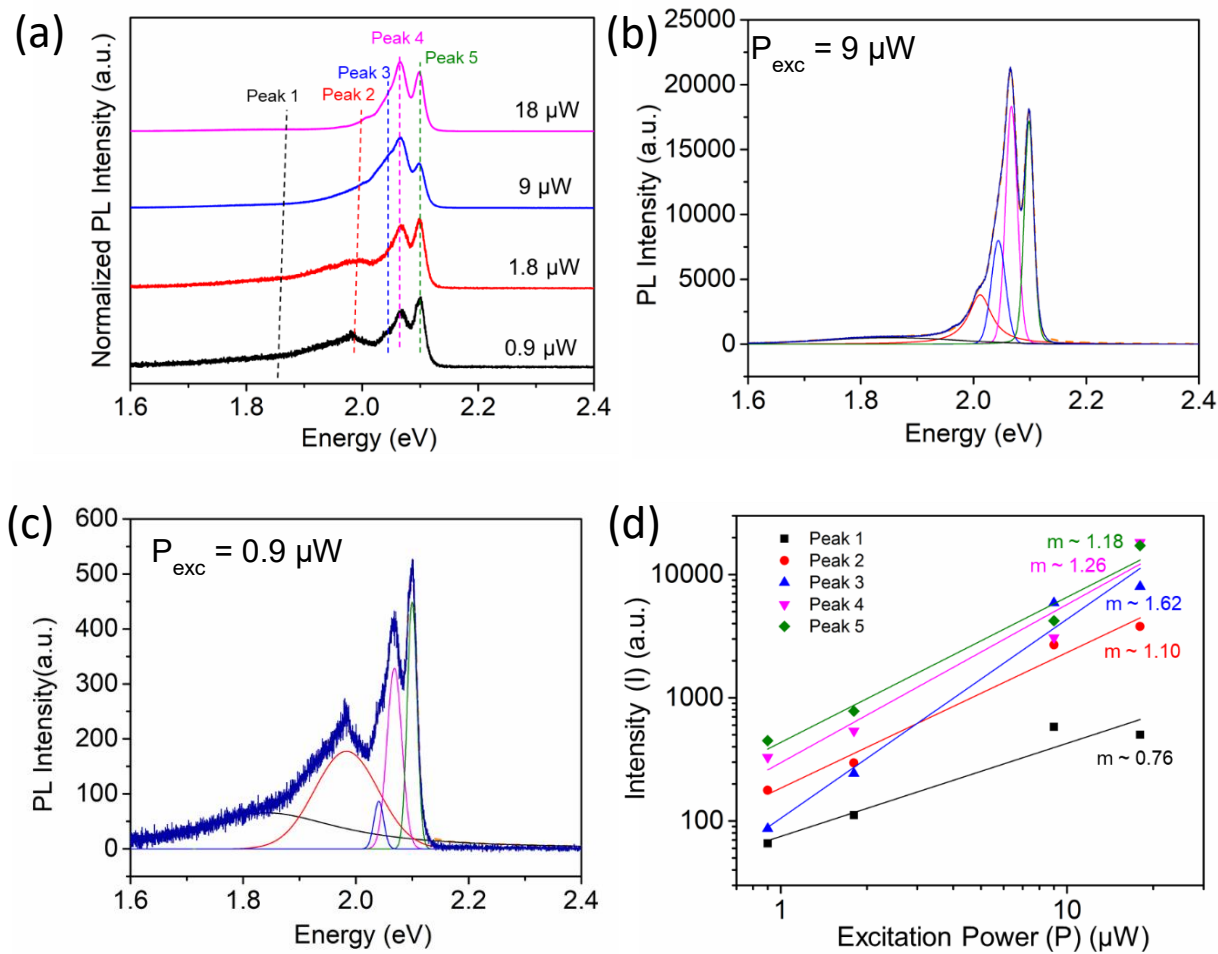

**Figure S8.** PL spectra of transferred WS$_2$ at different laser power (b) Fitting of the PL spectra acquired with laser power of 9 µW, (c) Fitting of the PL spectra acquired with laser power of 0.9 µW and (d) Plot of peak intensity as a function of laser power.

Solid lines are fittings based on the power dependence of PL peaks: $I \sim P^m$. Here, logarithmic scales are used for plotting, so that the slope of linear fitting represents the numeric power (m). Peak 1 has a sublinear power dependence (m ~ 0.76), which is a typical characteristic of defect-bound excitons. Peak 3 has a significant superlinear power dependence (m ~ 1.62), which can be attributed to biexciton emission. The two peaks with highest energies (peak 4 and peak 5) are attributed to negative trion ($X^-$) and exciton ($X^0$), respectively. It should be noted that peak 2 has a power dependence of m ~ 1.10, which is not likely related to defect-induced emissions. The origin of peak 2 needs to be understood in our future works by lower temperature PL measurements (down to 4 K), as well as power-dependent PL studies with a larger excitation power range.

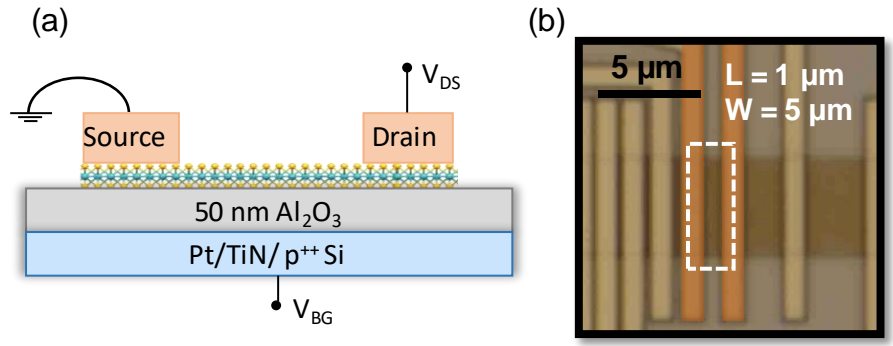

**Figure S9.** (a) Schematic of the WS$_2$ transistor. (b) Optical image of the device used for measurements in the manuscript.

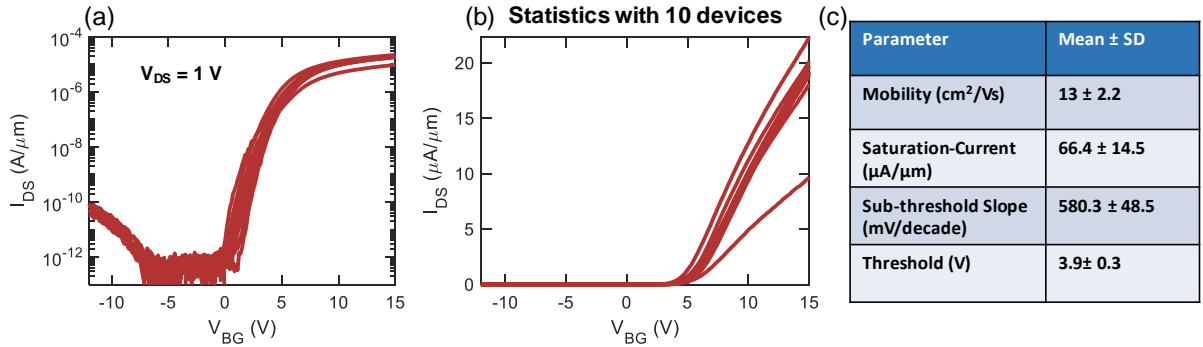

**Figure S10**. Transfer characteristics of 10 devices measured across the wafer in (a) log scale and, (b) linear scale. (c) Extracted mean and standard deviation for various device parameters such as the mobility, saturation-current, sub-threshold slope and threshold voltage. Minimal device-to-device variation indicates uniform and high quality WS2 growth across the wafer.

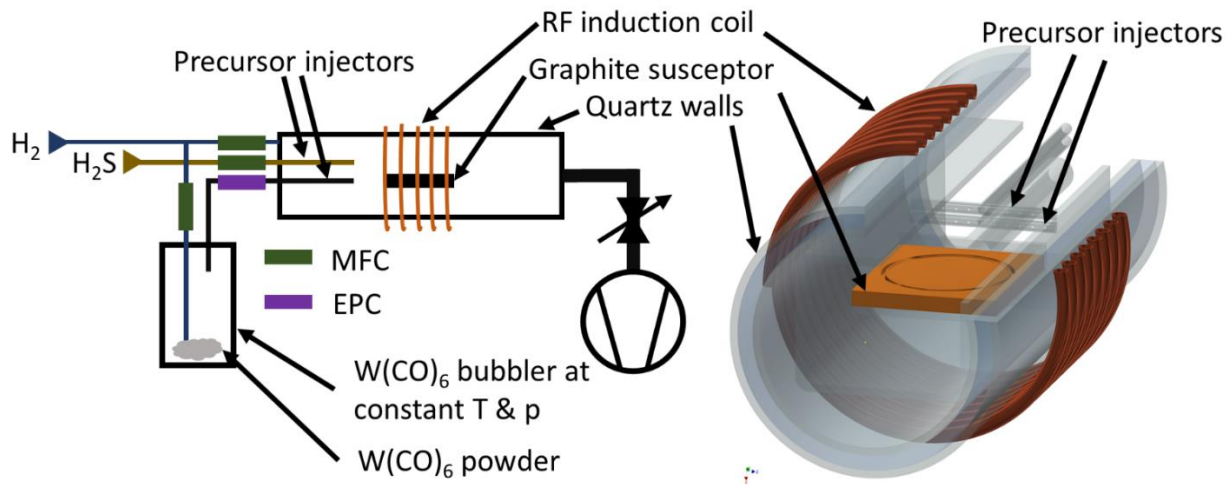

**Figure S11.** Simplified illustration of the MOCVD system (left) and reactor (right) used for the growth of $WS_2$ films.

**Synthesis description.** Simplified schematics of the MOCVD system and reactor are shown in Figure S6 and described in more detail in reference [2]. The MOCVD reactor consists of a horizontal water-cooled quartz tube with an inductively heated SiC-coated graphite susceptor that can hold a 2" wafer and includes gas foil rotation to improve the uniformity of the films. Precursor gases are delivered through separate run/vent manifolds and inlet tubes and are allowed to mix approximately 7 cm before the graphite susceptor. Loading and unloading of samples occurs through a $N_2$-ventilated glovebox to protect the reactor and samples from exposure to atmospheric oxygen and water.

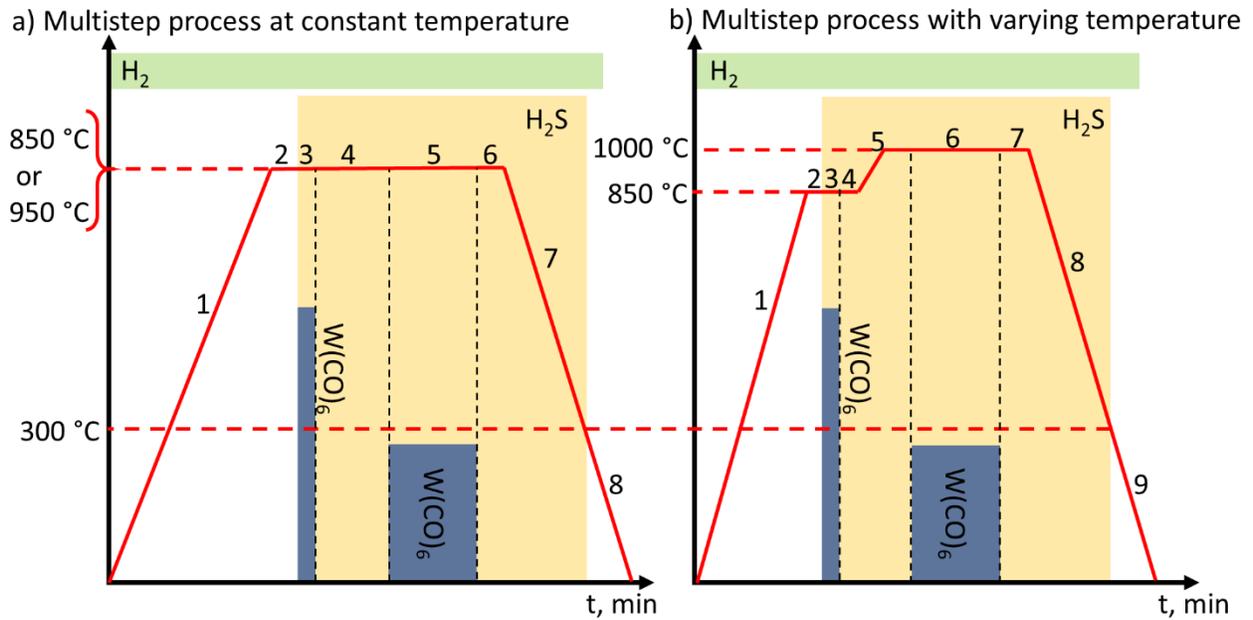

**Figure S12.** Illustration of precursor and carrier gas flow rates and substrate temperatures during the multistep process at a) constant and b) variable temperature.

A multi-step growth process (Figure S7) was used for MOCVD growth of the WS$_2$ films on sapphire. First step (1) is heat up to a growth temperature of 850 °C or 950 °C at constant pressure under a flow of H$_2$. Second step (2) is substrate annealing in H$_2$ for 10 min. Step (3) is nucleation of WS$_2$ when H$_2$S and W(CO)$_6$ are introduced into the reactor. Step (4) is a ripening step under H$_2$S and H$_2$ flow while W(CO)$_6$ is removed from the reactor. This is followed by growth step (5) where W(CO)$_6$ is re-introduced into the reactor at half the flow rate used for nucleation; H$_2$S is kept constant. After the growth, in step (6) the sample is annealed at growth temperature for 10 min under the flow of H$_2$S and H$_2$. This is followed by cooldown step (7) under H$_2$S and H$_2$ flow. H$_2$S is removed from the reactor at 300 °C and N$_2$ pump-purge cycle is initiated at 200 °C in step (8). Figure S2(b) is an illustration of gas flows and temperature of the multistep process employing variable temperatures. This process is in general similar to the process at constant temperature with the difference being that initial nucleation is performed at 850 °C and during the ripening step, the temperature is increased to 1000 °C (5) where growth (6) and postgrowth annealing (7) is performed. Timing of all the steps is the same.